\documentclass[twoside]{article}
\usepackage[accepted]{aistats2018}

\usepackage{hyperref}
\usepackage[labelfont=bf]{caption}
\usepackage[numbers]{natbib}
\usepackage[page]{appendix}
\usepackage[subrefformat=parens]{subcaption}
\usepackage[dvipsnames]{xcolor}

\usepackage{amsfonts}
\usepackage{amsmath}
\usepackage{amssymb}
\usepackage{bm}
\usepackage{booktabs}
\usepackage{enumitem}
\usepackage{framed}
\usepackage{graphicx}
\usepackage{lipsum}
\usepackage{mathtools}
\usepackage{placeins}
\usepackage{sidecap}
\usepackage{standalone}
\usepackage{textcomp}
\usepackage{tikz}

\bibpunct[; ]{[}{]}{,}{n}{,}{,}

\usetikzlibrary{arrows}
\usetikzlibrary{bayesnet}
\usetikzlibrary{calc}
\usetikzlibrary{decorations.pathreplacing}
\usetikzlibrary{patterns}
\usetikzlibrary{positioning}

\hypersetup{colorlinks}
\hypersetup{linkcolor={red!50!black}}
\hypersetup{citecolor={MidnightBlue}}
\newcommand{\bc}{\mathbf{c}}
\newcommand{\x}{\mathbf{x}}
\newcommand{\y}{\mathbf{y}}
\newcommand{\z}{\mathbf{z}}

\newcommand{\crp}{\textrm{CRP}}
\newcommand{\set}[1]{\left\lbrace{#1}\right\rbrace}

\newcommand{\commentsize}{\footnotesize}
\newcommand{\mathcomment}[1]{\textrm{\commentsize #1}}

\makeatletter
\newcommand*{\defeq}{
  \mathrel{
    \rlap{%
      \raisebox{0.3ex}{$\m@th\cdot$}}%
      \raisebox{-0.3ex}{$\m@th\cdot$}}%
  =}
\makeatother

\newcommand\ttag{\addtocounter{equation}{1}\tag{\theequation}}

\newcounter{algori}

\begin{document}

\runningtitle{Temporally-Reweighted Chinese Restaurant Process Mixtures
  for Multivariate Time Series}

\runningauthor{Saad and Mansinghka}

\twocolumn[

  \aistatstitle{Temporally-Reweighted Chinese Restaurant Process Mixtures for \\
    Clustering, Imputing, and Forecasting Multivariate Time Series}

  \aistatsauthor{Feras A. Saad \And Vikash K. Mansinghka}

  \aistatsaddress{
    Probabilistic Computing Project \\
    Massachusetts Institute of Technology
    \And Probabilistic Computing Project \\
    Massachusetts Institute of Technology
    }
]

\begin{abstract}
This article proposes a Bayesian nonparametric method for forecasting,
imputation, and clustering in sparsely observed, multivariate time series data.
The method is appropriate for jointly modeling hundreds of time series with
widely varying, non-stationary dynamics. Given a collection of $N$ time series,
the Bayesian model first partitions them into independent clusters using a
Chinese restaurant process prior. Within a cluster, all time series are modeled
jointly using a novel ``temporally-reweighted'' extension of the Chinese
restaurant process mixture. Markov chain Monte Carlo techniques are used to
obtain samples from the posterior distribution, which are then used to form
predictive inferences. We apply the technique to challenging forecasting and
imputation tasks using seasonal flu data from the US Center for Disease Control
and Prevention, demonstrating superior forecasting accuracy and competitive
imputation accuracy as compared to multiple widely used baselines. We further
show that the model discovers interpretable clusters in datasets with hundreds
of time series, using macroeconomic data from the Gapminder Foundation.
\end{abstract}

\section{Introduction}
\label{sec:introduction}

Multivariate time series data is ubiquitous, arising in domains such as
macroeconomics, neuroscience, and public health. Unfortunately, forecasting,
imputation, and clustering problems can be difficult to solve when there are
tens or hundreds of time series. One challenge in these settings is
that the data may reflect underlying processes with widely varying,
non-stationary dynamics \citep{fulcher2014}.
Another challenge is that standard parametric approaches such as state-space
models and vector autoregression often become statistically and numerically
unstable in high dimensions \citep{koop2013}. Models from these families further
require users to perform significant custom modeling on a per-dataset basis, or
to search over a large set of possible parameter settings and model
configurations. In econometrics and finance, there is an increasing need for
multivariate methods that exploit sparsity, are computationally efficient, and
can accurately model hundreds of time series (see introduction of
\citep{gruber207}, and references therein).

This paper presents a nonparametric Bayesian method for multivariate time series
that aims to address some of the above challenges. The model is based on two
extensions to Dirichlet process mixtures. First, we introduce a recurrent
version of the Chinese restaurant process mixture to capture temporal
dependences. Second, we add a hierarchical prior to discover groups of time
series whose underlying dynamics are modeled jointly. Unlike autoregressive
models, our approach is designed to interpolate in regimes where it has seen
similar history before, and reverts to a broad prior in previously unseen
regimes. This approach does not sacrifice predictive accuracy, when there is
sufficient signal to make a forecast or impute missing data.

We apply the method to forecasting flu rates in 10 US regions using flu,
weather, and Twitter data from the US Center for Disease Control and Prevention.
Quantitative results show that the method outperforms several Bayesian and
non-Bayesian baselines, including Facebook Prophet, multi-output Gaussian
processes, seasonal ARIMA, and the HDP-HMM. We also show competitive imputation
accuracy with widely used statistical techniques. Finally, we apply the method
to clustering hundreds of macroeconomic time series from Gapminder, detecting
meaningful clusters of countries whose data exhibit coherent temporal patterns.

\section{Related Work}
\label{sec:related-work}

The temporally-reweighted Chinese restaurant process (TRCRP) mixture we
introduce in Section~\ref{sec:temporally-reweighted-crp} can be directly seen as
a time series extension to a family of nonparametric Bayesian regression models
for cross-sectional data
\citep{ishwaran2003,shahbaba2009jmlr,park2010,muller2010,muller2011}.
These methods operate on an exchangeable data sequence $\set{x_i}$ with
exogenous covariates $\set{\y_i}$; the prior CRP cluster probability $p(z_i=k)$
for each observation $x_i$ is reweighted based on $\mathbf{y}_i$.
Our method extends this idea to a time series $\set{x_t}$; the prior CRP cluster
probability $p(z_t=k)$ for $x_t$ is now reweighted based on the $p$ previous
values $\x_{t-1:t-p}$.
Moreover, the hierarchical extension in
Section~\ref{subsec:generative-model-structure} coincides with CrossCat
\citep{mansinghka2016}, when all temporal dependencies are removed (by setting
$p=0$).

Temporal extensions to the Dirichlet process have been previously used in the
context of dynamic clustering \citep{zhu2005,ahmed2008dynamic}. The latter work
derives a recurrent CRP as the limit of a finite dynamic mixture model. Unlike
the method in this paper, those models are used for clustering batched data and
dynamic topic modeling \citep{blei2006}, rather than data analysis tasks such as
forecasting or imputation in real-valued, multivariate time series.

For multivariate time series, recent nonparametric Bayesian methods include
using the dependent Dirichlet process for dynamic density estimation
\citep{rodriguez2008}; hierarchical DP priors over the state in hidden Markov
models \citep[HDP-HMM;][]{fox2009,johnson2013}; Pitman-Yor mixtures of
non-linear state-space models for clustering \citep{barajas2014}; and DP
mixtures \citep{caron2007} and Polya trees \citep{barajas2016} for modeling
noise distributions.
As nonparametric Bayesian extensions of state-space models, all of these
approaches specify priors that fall under distinct model classes to the one
developed in this paper.
They typically encode parametric assumptions (such as linear autoregression and
hidden-state transition matrices), or integrate explicit specifications of
underlying temporal dynamics such as seasonality, trends, and time-varying
functionals.
Our method instead builds purely empirical models and uses simple infinite
mixtures to detect patterns in the data, without relying on dataset-specific
customizations.
As a multivariate interpolator, the TRCRP mixture is best applied to time series
where there is no structural theory of the temporal dynamics, and where there is
sufficient statistical signal in the history of the time series to inform
probable future values.

To the best of our knowledge, this paper presents the first multivariate,
nonparametric Bayesian model that provides strong baseline results without
specifying custom dynamics on a problem-specific basis; and that has been
benchmarked against multiple Bayesian and non-Bayesian techniques to cluster,
impute, and forecast sparsely observed real-world time series data.

\section{Temporally-Reweighted Chinese Restaurant Process Mixture Model}
\label{sec:temporally-reweighted-crp}

We first outline the notations and basic setup assumed throughout this paper.
Let $\set{\x^n: n = 1,\dots,N}$ denote a collection of $N$ discrete-time series,
where the first $T$ variables of the $n$\textsuperscript{th} time series is
$\x^n_{1:T} = (x^n_1, x^n_2, \dots, x^n_T)$. Slice notation is used to index
subsequences of variables, so that $\x^n_{t_1:t_2} = (x^n_{t_1}, \dots,
x^n_{t_2})$ for $t_1 < t_2$. Superscript $n$ will be often be omitted when
discussing a single time series. The remainder of this section develops a
generative process for the joint distribution of all random variables
$\set{x_t^n: t=1,\dots,N, n = 1,\dots,N}$ in the $N$ time series, which we
proceed to describe in stages.

\subsection{Background: CRP representation of Dirichlet
process mixture models}
\label{subec:background-crp-mixture}

Our approach is based on a temporal extension of the standard Dirichlet process
mixture (DPM), which we review briefly.
First consider the standard DPM in the non-temporal setting \citep{escobar1995},
with concentration $\alpha$ and base measure $\pi_\Theta$.
The joint distribution of a sequence of $m$ exchangeable random variables $(x_1,
\dots, x_m)$ is:
\begin{align*}
P \sim DP(\alpha, \pi_\Theta),
\quad \theta^*_j \mid P \sim P,
\quad x_j \mid \theta^*_j \sim F(\cdot \mid \theta^*_j).
\end{align*}
The DPM can be represented in terms of the Chinese restaurant process
\citep{aldous1985}.
As $P$ is almost-surely discrete, the $m$ draws $\set{\theta^*_j} \sim P$
contain repeated values, thereby inducing a clustering among data $x_j$.
Let $\lambda_F$ be the hyperparameters of $\pi_\Theta$, $\set{\theta_k}$ be the
unique values among the $\set{\theta^*_j}$, and $z_j$ denote the cluster
assignment of $x_j$ which satisfies $\theta^*_j = \theta_{z_j}$.
Define $n_{jk}$ to be the number of observations $x_{i}$ with $z_i=k$ for $i<j$.
Using the conditional distribution of $z_j$ given previous cluster assignments
$\z_{1:j-1}$, the joint distribution of exchangeable data sequence
$(x_1,x_2,\dots)$ in the CRP mixture model can be described sequentially:
\begin{align}
&\set {\theta_k} \overset{\mathrm{iid}}{\sim}
  \pi_\Theta(\cdot \mid \lambda_F) \notag \\[7pt]
&\begin{aligned}
&\Pr\left[z_j = k \mid \z_{1:j-1}; \alpha\right] && (j=1,2,\dots)  \\
&\quad \propto \begin{cases}
  n_{jk}  & \mathcomment{if } 1 \le k \le \max{(\z_{1:j-1})} \\
  \alpha  & \mathcomment{if } k = \max{(\z_{1:j-1})} + 1
  \end{cases}
\end{aligned} \label{eq:crp-probability} \\
&x_j \mid z_j, \set{\theta_k} \sim F(\cdot|\theta_{z_j}) \notag
\end{align}
The CRP mixture model \eqref{eq:crp-probability}, and algorithms for posterior
inference, have been studied extensively for nonparametric modeling in a variety
of statistical applications (for a survey see \citep{teh2011}, and references
therein).


\begin{figure}[ht]
\centering

\begin{tikzpicture}[transform shape, scale=0.9]

\tikzstyle{constant}=[ fill=black];


\node[constant, label={above:$\alpha, \lambda_G$},] (alpha-lambda) {};


\node[latent, below left = 1 and 2 of alpha-lambda,] (z_1) {$z_1$};
\node[latent, right = 1 of z_1] (z_2) {$z_2$};

\node[latent, below right = 1 and 2 of alpha-lambda,] (z_4) {$z_4$};
\node[latent, left = 1 of z_4] (z_3) {$z_3$};
\node[const, right = .2 of z_4] (z_dots) {$\cdots$};


\node[obs, below = of z_1] (x_1) {$x_1$};
\node[obs, below = of z_2] (x_2) {$x_2$};
\node[obs, below = of z_3] (x_3) {$x_3$};
\node[obs, below = of z_4] (x_4) {$x_4$};
\node[const] at (x_4 -| z_dots) (x_dots) {$\cdots$};


\node[const, left = .5 of x_1] (x_0) {$x_{0}$};


\node[latent, left = .5 of x_0] (thetas) {$\theta_k$};
\node[draw, minimum height = 50, minimum width = 40] at (thetas) (plate) {};
\node[anchor=south] at (plate.south) {\scriptsize $k{=}1,2,\dots$};

\draw[-latex] (thetas.330) to[bend right = 15] (x_1.260);
\draw[-latex] (thetas.330) to[bend right = 15] (x_2.260);
\draw[-latex] (thetas.330) to[bend right = 15] (x_3.260);
\draw[-latex] (thetas.330) to[bend right = 15] (x_4.260);

\node[constant, label={above:$\lambda_F$}] at (thetas |- z_1) (lambdaF) {};
\draw[-latex] (lambdaF) -- (thetas);


\draw[-latex] (alpha-lambda) -- (z_1.north);
\draw[-latex] (alpha-lambda) -- (z_2.north);
\draw[-latex] (alpha-lambda) -- (z_3.north);
\draw[-latex] (alpha-lambda) -- (z_4.north);

\draw[-latex] (z_1) to (z_2);
\draw[-latex] (z_1) to [bend right = 25] (z_3);
\draw[-latex] (z_1) to [bend right = 25] (z_4);

\draw[-latex] (z_2) to (z_3);
\draw[-latex] (z_2) to [bend right = 25] (z_4);
\draw[-latex] (z_3) to (z_4);

\draw[-latex] (z_1) -- (x_1);
\draw[-latex] (z_2) -- (x_2);
\draw[-latex] (z_3) -- (x_3);
\draw[-latex] (z_4) -- (x_4);

\draw[-latex] (x_1.45) -- (z_2);
\draw[-latex] (x_2.45) -- (z_3);
\draw[-latex] (x_3.45) -- (z_4);
\draw[-latex] (x_0.45) -- (z_1);
\end{tikzpicture}

\caption{Graphical model for the TRCRP mixture in a single time series $\x =
(x_1, x_2, \dots)$ with lagged window size $p = 1$.}
\label{fig:graphical-model}

\end{figure}
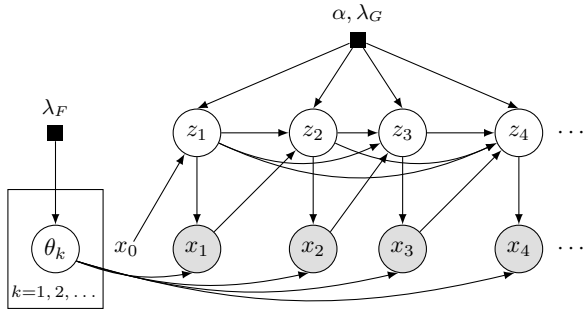

\subsection{The temporally-reweighted CRP mixture for modeling a single time
series}
\label{subsec:generative-model-single}


\begin{figure*}[ht]

\begin{subfigure}[b]{.5\linewidth}
\centering
\begin{flalign*}
&\mathcomment{\textbf{1}. Sample concentration parameter of CRP}
  \span \\
&\qquad\alpha \sim \textrm{Gamma(1,1)}
  \span \\
&\mathcomment{\textbf{2}. Sample model hyperparameters }
  & (n=1,2,\dots,N) \\
&\qquad \lambda_G^n \sim H^n_G
  & \\
&\qquad \lambda_F^n \sim H^n_F
  & \\
&\mathcomment{\textbf{3}. Sample distribution parameters of } F
  & (n=1,2,\dots,N) \\
&\qquad \theta^n_1,\theta^n_2,\dots
    \overset{\mathrm{iid}}{\sim} \pi_\Theta(\cdot \vert \lambda^n_F)
  & \\
&\mathcomment{\textbf{4}. Assume first $p$ values are known }
  & (n=1,2,\dots, N) \\
&\qquad \x^n_{-p+1:0} \defeq (x^n_{-p+1}, \dots, x^n_{0})
  & \\
&\mathcomment{\textbf{5}. Sample time series observations }
  & (t=1,2,\dots) \\
&\quad \mathcomment{\textbf{5.1} Sample temporal cluster assignment } z_t
  \span \\
&\qquad \begin{array}{@{}l}
  \Pr\left[z_t = k \mid \z_{1:t-1},
    \x^{1:N}_{t-p:t-1}, \alpha, \lambda^{1:N}_G\right] \\[2.5pt]
  \quad \propto {\crp}(k|\alpha, \z_{1:t-1})
    \prod_{n=1}^N G(\x^n_{t-p:t-1}; D^n_{tk}, \lambda_G^n) \\[2.5pt]
  \mathcomment{where }
    D^n_{tk} \defeq \set{\x^n_{t'-p:t'-1} \mid z_{t'} = k, 1 \le {t'} < t} \\
  \mathcomment{and }
    k=1,\dots, \max{(\z_{1:t-1})}+1
\end{array}  \span \\
&\quad \mathcomment{\textbf{5.2} Sample data } x^n_t
  & (n=1,2,\dots,N) \\
&\qquad x^n_t \mid z_t, \set{\theta^n_k} \sim F(\cdot|\theta^{n}_{z_t})
  &
\end{flalign*}
\caption{Generative process for the multivariate TRCRP mixture}
\label{subfig:multivariate-generative-model-math}
\end{subfigure}\hfill
\begin{subfigure}[b]{.5\linewidth}
\includegraphics[width=\textwidth]{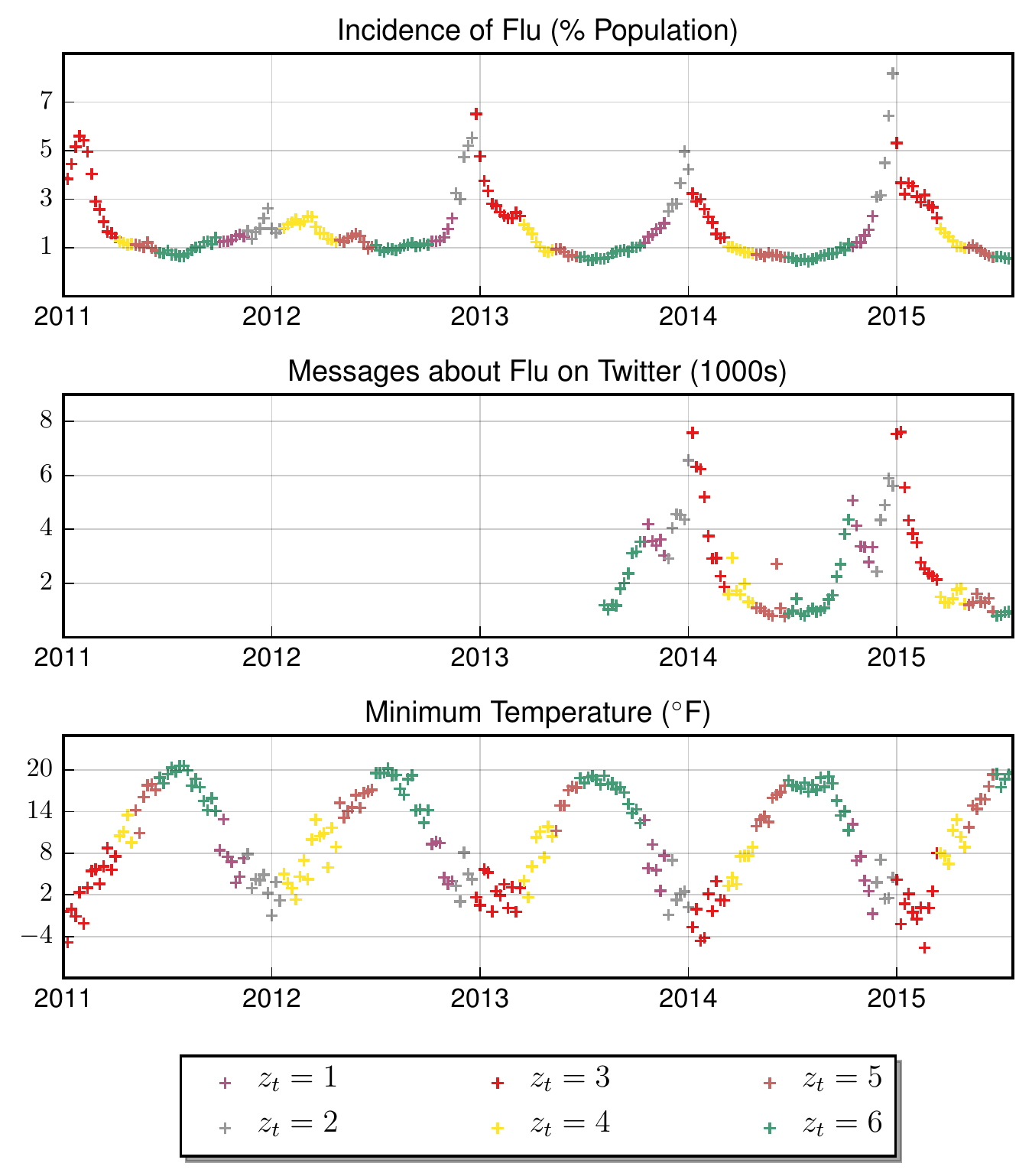}
\caption{Discovering flu season dynamics with the method}
\label{subfig:multivariate-generative-model-flu}
\end{subfigure}

\caption{\subref{subfig:multivariate-generative-model-math} Generative
model describing the joint distribution of $N$ dependent time series
$\set{\x^n}$ in the multivariate temporally-reweighted CRP mixture. Lagged
values for all time series are used for reweighting the CRP by $G$ in step
5.1.
Dependencies between time series are mediated by the shared temporal regime
assignment $z_t$, which ensures that all the time series have the same
segmentation of the time course into the different temporal regimes.
\subref{subfig:multivariate-generative-model-flu} Applying the TRCRP mixture
with $p=10$ weeks to model $\x^{\mathcomment{flu}}$, $\x^{\mathcomment{tweet}}$,
and $\x^{\mathcomment{temp}}$ in US Region 4. Six regimes describing the
seasonal behavior shared among the three time series are detected in this
posterior sample. Purple, gray, and red are the pre-peak rise, peak, and
post-peak decline during the flu season; and yellow, brown, and green represent
the rebound in between successive seasons. In 2012, the model reports no red
post-peak regime, reflecting the season's mild flu peak. See
Section~\ref{sec:applications} for quantitative experiments.}
\label{fig:multivariate-generative-model}

\end{figure*}

Our objective is to define a CRP-like process for a non-exchangeable
discrete-time series $(x_1, x_2, \dots)$, where there is now a temporal ordering
and a temporal dependence among the variables.
Instead of having $(x_t,z_t)$ be conditionally independent of all other data
given $\z_{1:t-1}$ as in the CRP mixture \eqref{eq:crp-probability}, we instead
consider using previous observations $\x_{1:t-1}$ when simulating $z_t$.
The main idea in our approach is to modify the CRP prior by having the cluster
probability $\Pr[z_t = k \mid \z_{1:t-1}]$ at step $t$ additionally account
for (i) the $p$ most recent observations $\x_{t-p:t-1}$, and (ii) collection of
lagged values $D_{tk} := \set{\x_{t'-p:t'-1} \mid z_{t'} = k, 1 \le t' < t}$ of
earlier data points $x_{t'}$ assigned to cluster $k$.
The distribution of time series $(x_1,x_2,\dots)$ in the
temporally-reweighted CRP (TRCRP) mixture is therefore:
\begin{flalign}
&{\set{\theta_k}} \overset{\mathrm{iid}}{\sim} \pi_\Theta(\cdot \mid \lambda_F)
  \notag \\[7pt]
&{\Pr}\left[z_t = k \mid \z_{1:t-1}, \x_{t-p:t-1}; \alpha, \lambda_G \right]
  & & (t=1,2,\dots) \notag \\
&\, \propto
  \rlap{$\begin{cases}
    n_{tk} \, G(\x_{t-p:t-1}; D_{tk}, \lambda_G)
      & \mathcomment{if } 1 \le k \le \max{(\z_{1:t-1})} \\
    \alpha \, G(\x_{t-p:t-1}; \lambda_G)
      & \mathcomment{if } k = \max{(\z_{1:t-1})} + 1
  \end{cases}$} \notag \\
&x_t \mid z_t, \set{\theta_k} \sim F(\cdot|\theta_{z_t}) & &
\label{eq:crp-probability-reweighted}
\end{flalign}
The main difference between the TRCRP mixture
\eqref{eq:crp-probability-reweighted} and the standard CRP mixture
\eqref{eq:crp-probability} is the term $G(\x_{t-p:t-1}; \lambda_G, D_{tk})$
which acts as a non-negative ``cohesion'' function $\mathbb{R}^p \to
\mathbb{R}^+$, parametrized by $D_{tk}$ and a bundle of real values $\lambda_G$.
This term measures how well the current lagged values $\x_{t-p:t-1}$ match the
collection of lagged values of earlier data $D_{tk}$ in each cluster $k$,
thereby introducing temporal dependence to the model.
The smoothness of the process depends on the choice of the window size $p$: if
$t_1$ and $t_2$ are close in time (relative to $p$) then they have overlapping
lagged values $\x_{t_1-p:t_1-1}$ and $\x_{t_2-p:t_2-1}$, so $G$ increases the
prior probability that $\set{z_{t_1} = z_{t_2}}$.
More generally, any pair of time points ${t_1}$ and ${t_2}$ that share similar
lagged values are a-priori more likely to have similar distributions for
generating $x_{t_1}$ and $x_{t_2}$, because $G$ increases the probability that
$\set{z_{t_1} = z_{t_2} = k}$, so that $x_{t_1}$ and $x_{t_2}$ are both drawn
from $F(\cdot|\theta_k)$.

Figure~\ref{fig:graphical-model} shows a graphical model for the TRCRP mixture
\eqref{eq:crp-probability-reweighted} with window size $p = 1$.
The model proceeds as follows: first assume the initial $p$ observations
$(x_{-p+1},\dots,x_{0})$ are fixed or have a known joint distribution.
At step $t$, the generative process samples a cluster assignment $z_t$, whose
probability of joining cluster $k$ is a product of (i) the CRP probability for
$\set{z_t=k}$ given all previous cluster assignments $\z_{1:t-1}$, and (ii) the
``cohesion'' term $G(\x_{t-p:t-1}; \lambda_G, D_{tk})$.
In Figure~\ref{fig:graphical-model}, edges between the $z_t$'s denote the CRP
probabilities, while edges from $x_{t-1}$ up to $z_t$ represent reweighting
the CRP by $G$. Cluster assignment $z_t$ identifies the temporal regime that
dictates the distribution of $x_t \sim F(\cdot\vert\theta_{z_t})$.
Observe that if $p=0$ or $G \propto 1$, then the model reduces to a standard CRP
mixture \eqref{eq:crp-probability} with no temporal dependence, since $(z_t,
x_t)$ are conditionally independent of the entire time series history
$\x_{1:t-1}$ given $\z_{1:t-1}$.
Also note that the model is not Markovian, due to the infinite coupling among
the latent $z_t$ (compare to the recurrent switching linear dynamical system of
\citep{barber2006}).

The data distribution $F$ in \eqref{eq:crp-probability-reweighted} is a Normal
distribution with Normal-InverseGamma prior $\pi_\Theta$:
\begin{align}
\pi_\Theta(\mu_k, \sigma^2_k \mid m, V, a, b)
  &= \mathrm{N}(\mu_k | m, \sigma_k^2V) \mathrm{IG}(\sigma_k^2 | a,b) \notag \\
F(x_t|\mu_k, \sigma_k)
  &= \mathrm{N}(x_t \mid \mu_{k}, \sigma^2_{k}), \label{eq:normal-inverse-gamma}
\end{align}
where $\theta_k = (\mu_k, \sigma_k^2)$ are the per-cluster parameters of $F$,
and $\lambda_F = (m,V,a,b)$ the hyperparameters of $\pi_\Theta$.
Conjugacy of $F$ and $\pi_\Theta$ \citep{bernardo1994} implies that $\theta_k$
can be marginalized out of the generative model
\eqref{eq:crp-probability-reweighted} (see Appendix~\ref{appx:mcmc-methods}).
As for $G$, it may in general be any non-negative weighting function which
assigns a high value to lagged data vectors that are ``similar'' to one
another.
Previous approaches Bayesian nonparametric regression constructed
covariate-dependent probability measures using kernel-based reweighting
\citep{dunson2007}.
Our method defines $G$ as a product of $p$ Student-T distributions whose
location, scale, and degrees of freedom depend on lagged data $D_{tk}$ in
cluster $k$:
\begin{gather}
\begin{split}
&G(\x_{t-p:t-1}; D_{tk}, \lambda_G)
= \prod_{i=1}^{p} G_i(x_{t-i} ; D_{tki}, \lambda_{Gi}) \\
&\; = \prod_{i=1}^{p} \mathrm{T}_{2a_{tki}} \left(
    x_{t-i} ; m_{tki}, b_{tki}\frac{1+V_{tki}}{a_{tki}}
    \right) \label{eq:reweighting-function-hypers}
\end{split}
\end{gather}
where hyperparameter $\lambda_{Gi} = (m_{i0}, V_{i0}, a_{i0}, V_{i0})$ and data
$D_{tki} = \set{x_{t'-i} : z_{t'} = k, 1 \le t' < t}$.
Equations for the data-dependent terms $(m_{tki}, V_{tki}, a_{tki}, b_{tki})$
are given in Appendix~\ref{appx:data-dependent-reweighting}.
We emphasize that $G$ itself is used for reweighting only; it does not define a
probability distribution over lagged data.
Mathematically, $G$ attracts $x_t$ towards a cluster $k$ that assigns
$\x_{t-p:t-1}$ a high density, under the posterior predictive of an axis-aligned
Gaussian having observed $D_{tk}$ \citep{murphy2007}.

\subsection{Extending the TRCRP mixture to multiple dependent time series}
\label{subsec:generative-model-multiple}


\begin{figure*}[ht]
\begin{tikzpicture}

\node[inner sep=0pt]
  (k-series-all)
  {\includegraphics[width=.24\textwidth]{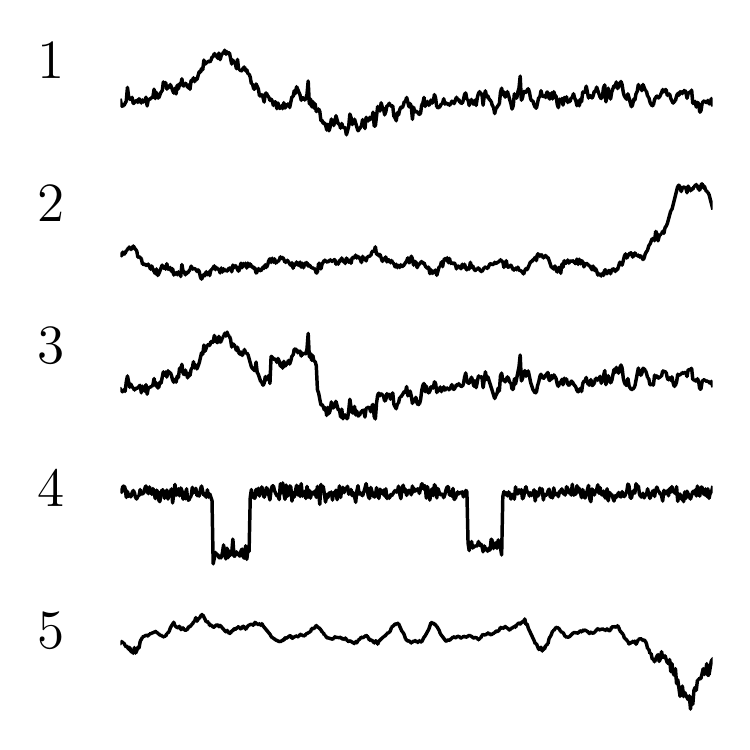}};

\node[inner sep=2pt, right=5.75 of k-series-all, anchor=west]
  (k-series-colored)
  {\includegraphics[width=.24\textwidth]{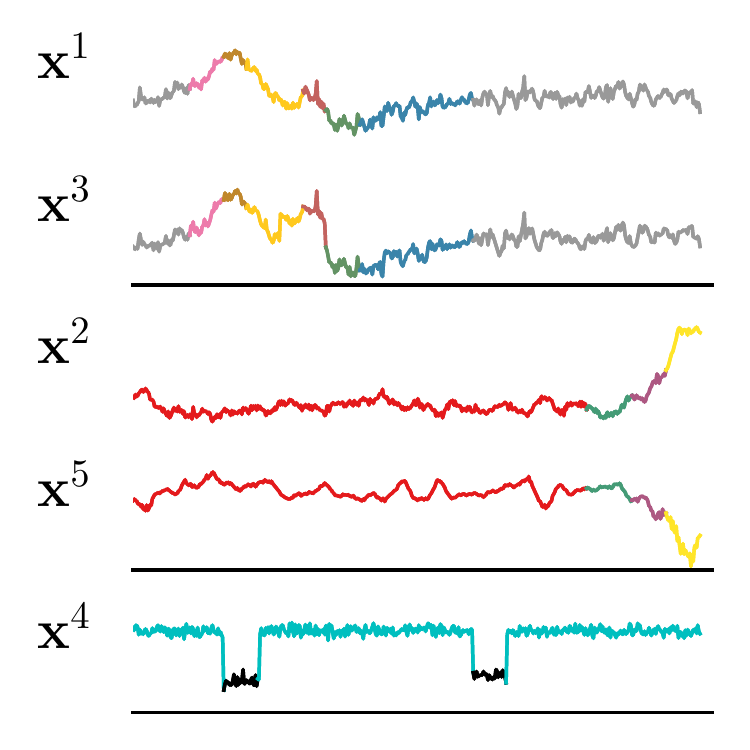}};

\node[
  right=.5cm of k-series-all.east,
  anchor=west,
  align=center,
] (crp-equation)
{\footnotesize $(c^1, c^2, c^3, c^4, c^5) \sim \crp$};

\node[
  circle,
  inner sep=2pt,
  draw=black,
  fill=black,
  align = center,
  right = 1.6 of crp-equation,
] (table-2) {};

\node[circle,
  inner sep=2pt,
  draw=black,
  fill=black,
  below=1 of table-2
] (table-3) {};

\node[
  circle,
  inner sep=2pt,
  draw=black,
  fill=black,
  above=1 of table-2,
] (table-1) {};

\draw[-latex] (k-series-all.35) -- (crp-equation.170);
\draw[-latex] (k-series-all.15) -- (crp-equation.175);
\draw[-latex] (k-series-all.0) -- (crp-equation.180);
\draw[-latex] (k-series-all.-15) -- (crp-equation.185);
\draw[-latex] (k-series-all.-35) -- (crp-equation.190);

\draw[-latex] (crp-equation.5) -- (table-1)
  node[pos=0.5, draw=black, inner sep=2pt, fill=white,
    minimum width=.9cm]{\scriptsize $c^1{=}c^3$};
\draw[-latex] (crp-equation.east) -- (table-2)
  node[pos=0.5, draw=black, inner sep=2pt, fill=white,
    minimum width=.9cm]{\scriptsize $c^2{=}c^5$};
\draw[-latex] (crp-equation.-5) -- (table-3)
  node[pos=0.5, draw=black, inner sep=2pt, fill=white,
    minimum width=.9cm]{\scriptsize $c^4$};;

\node[
  right=0 of k-series-colored.north east,
  anchor= north west,
  xshift=-.3cm,
  text width=1cm,
]
  {\begin{align*}
  &\sim \textrm{\footnotesize TRCRP Mixture} \\[30pt]
  &\sim \textrm{\footnotesize TRCRP Mixture} \\[20pt]
  &\sim \textrm{\footnotesize TRCRP Mixture}
  \end{align*}};

\node[below=0 of k-series-all] (label-original-series)
  {\footnotesize (a) Original $N=5$ time series};
\node[right= .3 of label-original-series, align=center] (label-crp)
  {\footnotesize (b) CRP over time series clusters};
\node[right= .3 of label-crp]  (label-clustered-series)
  {\footnotesize (c) TRCRP mixture within each cluster};
\end{tikzpicture}
\captionsetup{skip=5pt}
\caption{Hierarchical prior for learning the dependence structure between
multiple time series. Given $N$ EEG time series, we first nonparametrically
partition them by sampling an assignment vector $\bc^{1:N}$ from an ``outer''
CRP. Time series assigned to the same cluster are jointly generated using the
TRCRP mixture. Colored segments of each curve indicate the hidden states at each
time step (the shared latent variables within the cluster).}
\label{fig:generative-model-structure}
\hrule

\end{figure*}
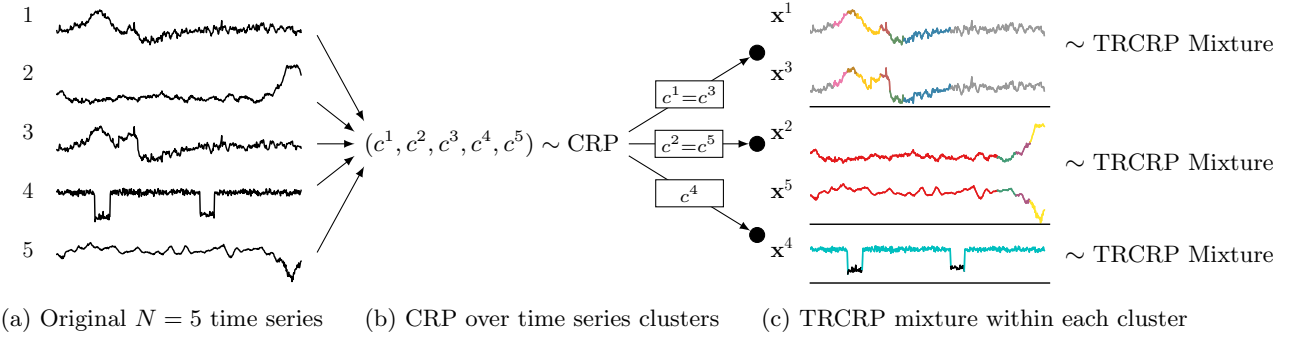

This section generalizes the univariate TRCRP mixture
\eqref{eq:crp-probability-reweighted} to handle a collection of $N$ time series
$\set{\x^n: n = 1,\dots,N}$, assumed for now to all be dependent.
At time $t$, we let the temporal regime assignment $z_t$ be shared among all the
time series, and use lagged values of all $N$ time series when reweighting the
CRP probabilities by the cohesion term $G$.
Figure~\ref{subfig:multivariate-generative-model-math} contains a step-by-step
description of the multivariate TRCRP mixture, with an illustrative
application in Figure~\ref{subfig:multivariate-generative-model-flu}.
It is informative to consider how $z_t$ mediates dependences between $\x^{1:N}$.
First, the model requires all time series to be in the same regime $z_t$ at time
$t$.
However, each time series has its own set of per-cluster parameters
$\set{\theta^n_k}$. Therefore, all the time series share the same segmentation
$\z_{1:T}$ of the time course into various temporal regimes, even though the
parametric distributions $F(\cdot\vert\theta^n_k), n = 1,\dots,N$ within each
temporal regime $k \in \z_{1:T}$ differ.
Second, the model makes the ``naive Bayes'' assumption that data
$\set{x^n_t}_{n=1}^N$ at time $t$ are independent given $z_t$, and that the
reweighting term $G$ in step 5.1 factors as a product.
This characteristic is essential for numerical stability of the method in high
dimensional and sparse regimes, while still maintaining the ability to recover
complex distributions due to the infinite CRP mixture.

\subsection{Learning the dependence structure between multiple time series}
\label{subsec:generative-model-structure}

The TRCRP mixture in Figure~\ref{subfig:multivariate-generative-model-math}
makes the restrictive assumption that all time series $\x^{1:N}$ are dependent
with one another.
However, with dozens or hundreds of time series whose temporal regimes are not
well-aligned, forcing a single segmentation sequence $\z_{1:T}$ to apply to all
$N$ time series will result in a poor fit to the data.
We relax this assumption by introducing a hierarchical prior that allows the
model to determine which subsets of the $N$ time series are probably
well-described by a joint TRCRP model.
The prior induces sparsity in the dependencies between the $N$ time series by
first nonparametrically partitioning them using an ``outer'' CRP. Within a
cluster, all time series are modeled jointly using the multivariate TRCRP
mixture described in Figure~\ref{subfig:multivariate-generative-model-math}:
\begin{align}
(c^1, c^2, \dots, c^N) &\sim \crp(\cdot | \alpha_0)
  \label{eq:crp-over-time-series} \\
\set{\x^n : c^n = k} &\sim \textrm{TRCRP Mixture} \notag \\
&\left(k = 1, \dots, \max{\bc^{1:N}}\right), \notag
\end{align}
where $c^n$ is the cluster assignment of $\x^n$.
Figure~\ref{fig:generative-model-structure} shows an example of this structure
learning prior applied to five EEG time series.
In the second cluster of panel (c), the final yellow segment illustrates two
time series sharing the latent regime at each time step, but having different
distributions within each regime.

\section{Posterior Inferences via Markov Chain Monte Carlo}
\label{sec:posterior-compuations}


\begin{figure*}[ht]

\begin{subfigure}{\linewidth}
\centering
\includegraphics[width=\linewidth]{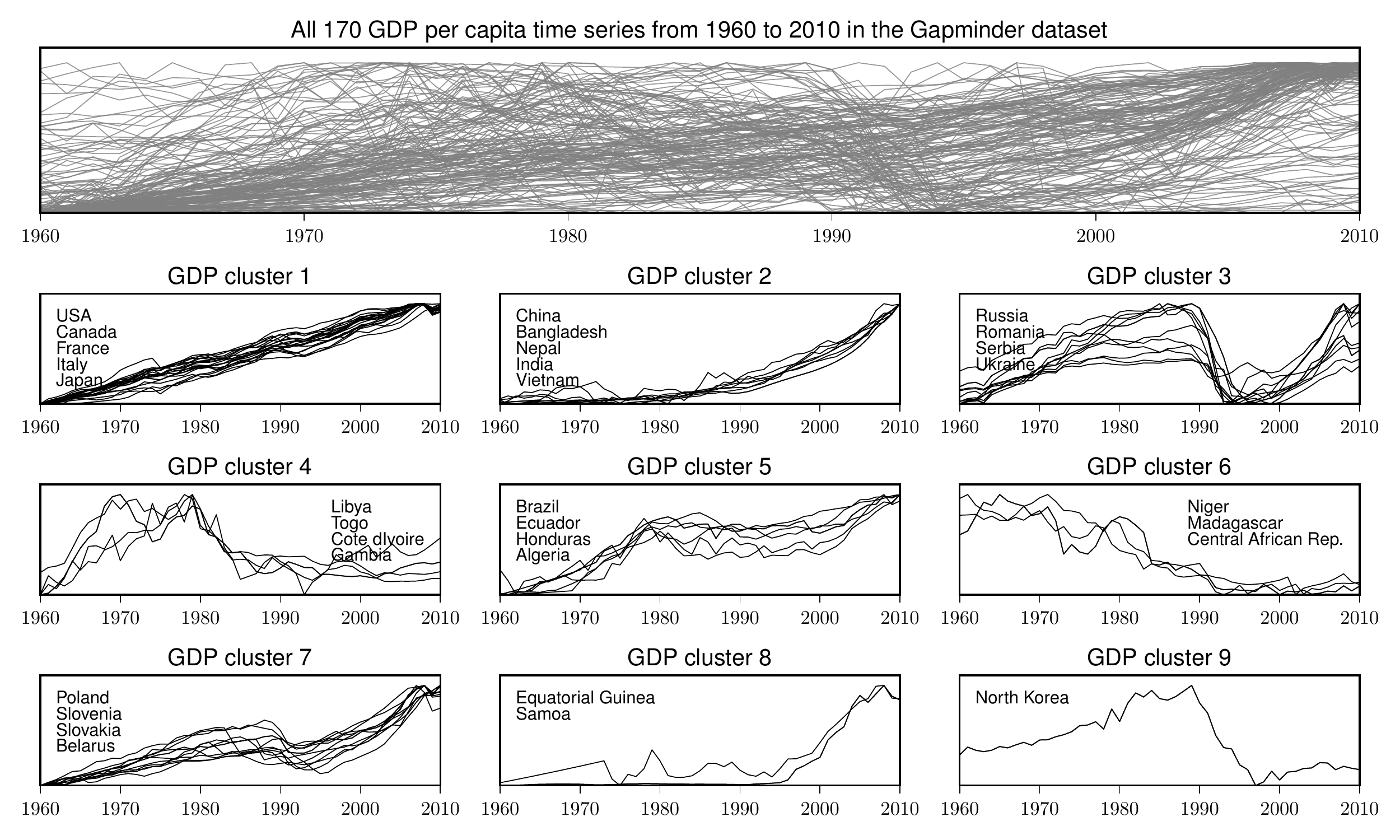}
\end{subfigure}
\captionsetup{skip=0pt}
\caption{Given GDP per capita data for 170 countries from 1960-2010, the
hierarchical TRCRP mixture \eqref{eq:crp-over-time-series} detects qualitatively
distinct temporal patterns. The top panel shows an overlay of all the time
series; nine representative clusters averaged over 60 posterior samples are
shown below. Countries within each cluster, of which a subset are labeled, share
similar political, economic, and/or geographic characteristics. For instance,
cluster 1 contains Western democracies with stable economic growth over 50 years
(slight dip in 2008 is the financial crash). Cluster 2 includes China and India,
whose GDP growth rates have outpaced those of industrialized nations since the
1990s. Cluster 3 contains former communist nations, whose economies tanked after
fall of the Soviet Union. Outliers such as Samoa, Equatorial Guinea, and North
Korea can be seen in clusters 8 and 9.}
\label{fig:cluster-datasets-gdp}

\end{figure*}


\begin{figure*}[ht]

\begin{subfigure}[t]{.7\textwidth}
\captionsetup{skip=0pt}
\subcaption{Four representative flu time series imputed jointly}
\label{subfig:imputation-baselines-imputations}
\includegraphics[width=\linewidth]{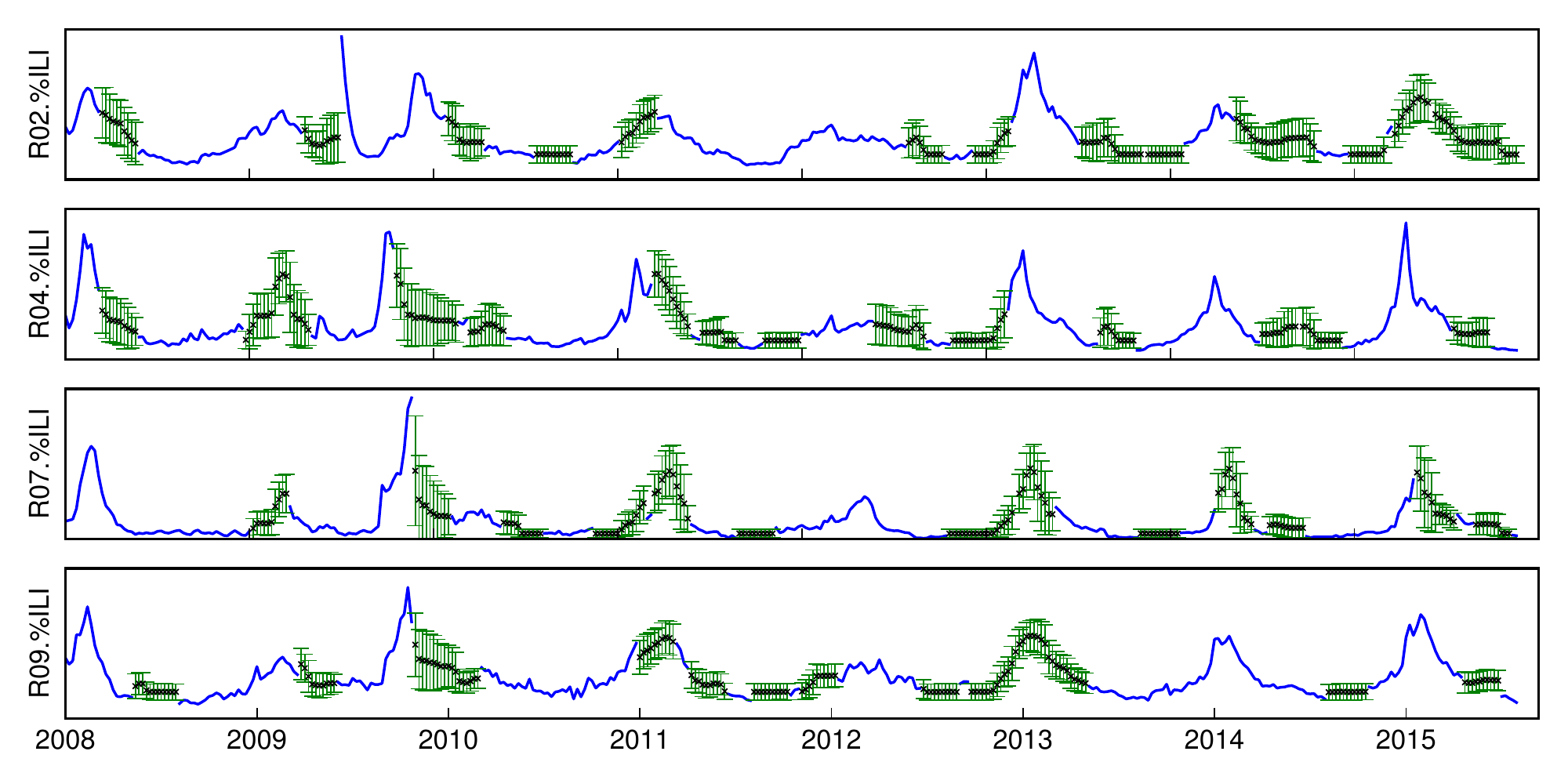}
\end{subfigure}
\begin{subfigure}[t]{.3\textwidth}
\captionsetup{skip=0pt}
\subcaption{Example imputations in R09}
\label{subfig:imputation-baselines-linear-nonlinear}
\includegraphics[width=\linewidth]{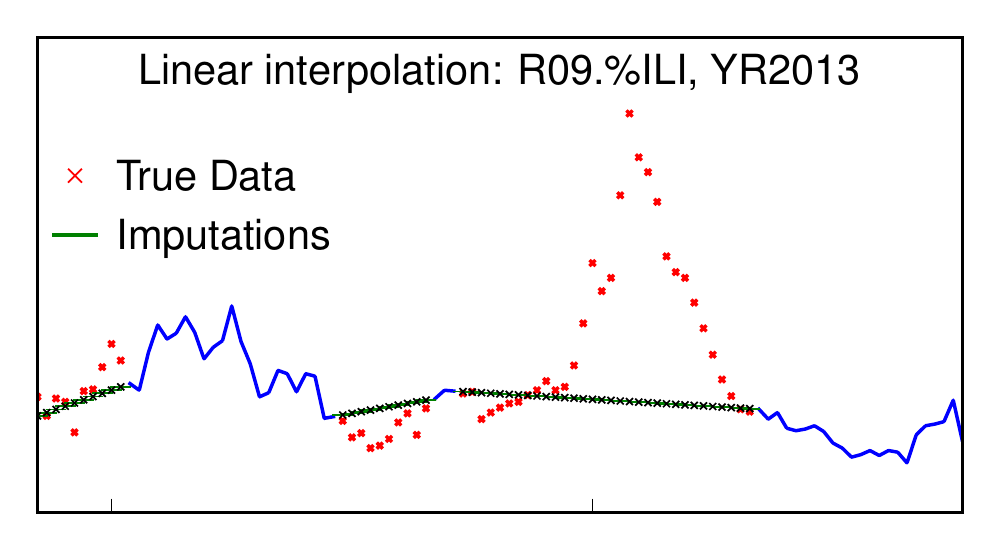}
\includegraphics[width=\linewidth]{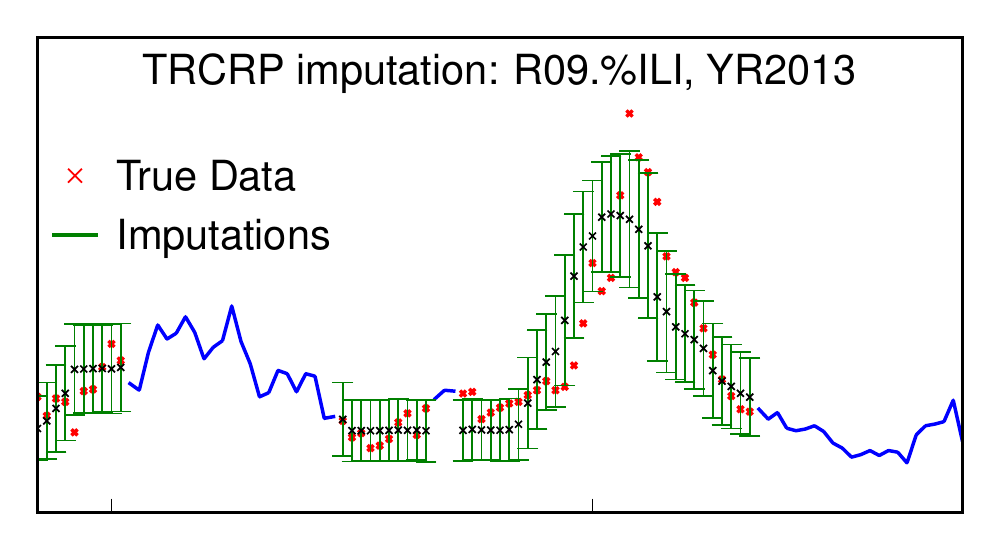}%
\end{subfigure}

\begin{subtable}{\linewidth}
\footnotesize
\setlength\tabcolsep{1.2pt}
\refstepcounter{subfigure}
\label{subfig:imputation-baselines-data}
\begin{tabular*}{\textwidth}{@{\extracolsep{\fill}}lllllllllll}
\toprule
\multicolumn{11}{c}{
  \textbf{\subref{subfig:imputation-baselines-data}}
  Mean absolute imputation errors in ten United States flu regions}
\\ \cmidrule(r){2-11}
\phantom
  & R01
  & R02
  & R03
  & R04
  & R05
  & R06
  & R07
  & R08
  & R09
  & R10
\\ \cmidrule(r){2-11}
Mean Imputation
  & $0.65_{(0.04)}$
  & $0.85_{(0.11)}$
  & $0.91_{(0.04)}$
  & $1.07_{(0.06)}$
  & $0.66_{(0.04)}$
  & $1.20_{(0.08)}$
  & $1.17_{(0.10)}$
  & $0.75_{(0.04)}$
  & $0.80_{(0.05)}$
  & $1.10_{(0.10)}$
\\
Linear Interpolation
  & $0.43_{(0.07)}$
  & $0.63_{(0.08)}$
  & $0.57_{(0.06)}$
  & $\mathbf{0.42}_{(0.04)}$
  & $0.44_{(0.04)}$
  & $0.71_{(0.08)}$
  & $0.71_{(0.09)}$
  & $0.35_{(0.03)}$
  & $0.43_{(0.05)}$
  & $0.72_{(0.06)}$
\\
Cubic Splines
  & $1.01_{(0.15)}$
  & $0.72_{(0.09)}$
  & $0.61_{(0.06)}$
  & $0.89_{(0.10)}$
  & $0.69_{(0.06)}$
  & $1.68_{(0.21)}$
  & $1.42_{(0.22)}$
  & $0.63_{(0.05)}$
  & $0.99_{(0.22)}$
  & $1.47_{(0.13)}$
\\
Multi-output GP
  & $0.36_{(0.04)}$
  & $0.57_{(0.11)}$
  & $0.32_{(0.02)}$
  & $0.58_{(0.07)}$
  & $0.30_{(0.03)}$
  & $0.57_{(0.04)}$
  & $0.62_{(0.04)}$
  & $\mathbf{0.34}_{(0.03)}$
  & $0.43_{(0.04)}$
  & $0.56_{(0.04)}$
\\
Amelia II
  & $0.29_{(0.03)}$
  & $0.52_{(0.11)}$
  & $0.25_{(0.02)}$
  & $0.45_{(0.03)}$
  & $\mathbf{0.29}_{(0.03)}$
  & $\mathbf{0.53}_{(0.04)}$
  & $\mathbf{0.53}_{(0.05)}$
  & $0.37_{(0.03)}$
  & $0.39_{(0.04)}$
  & $\mathbf{0.51}_{(0.03)}$
\\
TRCRP Mixture
  & $\mathbf{0.23}_{(0.03)}$
  & $\mathbf{0.47}_{(0.09)}$
  & $\mathbf{0.23}_{(0.02)}$
  & $0.49_{(0.04)}$
  & $0.31_{(0.03)}$
  & $0.55_{(0.05)}$
  & $0.75_{(0.07)}$
  & $\mathbf{0.34}_{(0.03)}$
  & $\mathbf{0.37}_{(0.03)}$
  & $0.67_{(0.07)}$
\\ \bottomrule
\end{tabular*}
\end{subtable}

\caption{Jointly imputing missing data in ten flu populations over eight
seasons.
\subref{subfig:imputation-baselines-imputations} Imputations and standard errors
in four of the time series. The TRCRP mixture accurately captures both
seasonal behavior as well as non-recurrent characteristics, such as the very
mild flu season in 2012.
\subref{subfig:imputation-baselines-data}
Comparing imputation quality with several baseline methods.
The TRCRP mixture ($p=10$ weeks) achieves comparable performance to Amelia II.
Cubic splines are completely ineffective due to long sequences without any
observations.
\subref{subfig:imputation-baselines-linear-nonlinear} While linear interpolation
may seem to be a good performer given its simplicity and mean errors, unlike the
TRCRP it cannot predict non-linear behavior when an entire flu season is
unobserved and entirely misses seasonality.}
\label{fig:imputation-baselines}

\end{figure*}


\begin{figure*}[ht]

\begin{subfigure}{\linewidth}
\includegraphics[width=\linewidth]{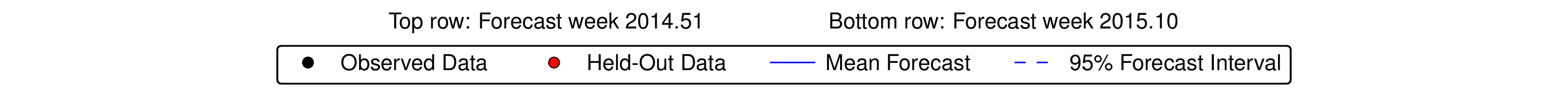}
\includegraphics[width=\linewidth]{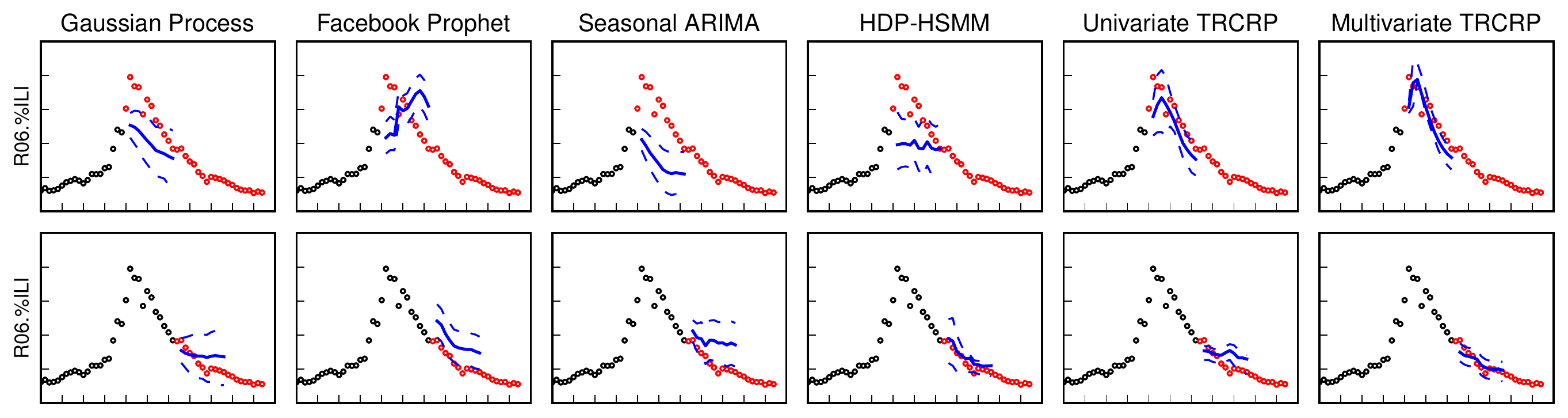}
\end{subfigure}
\begin{subtable}{\linewidth}
\footnotesize
\setlength\tabcolsep{1pt}
\begin{tabular*}{\textwidth}{@{\extracolsep{\fill}}lllllllllll}
\toprule
\multicolumn{11}{c}{
  Mean absolute flu prediction error for 10 forecast horizons (in weeks)
  averaged over 10 United States flu regions}
\\ \midrule
\phantom
  & $h=1$
  & $h=2$
  & $h=3$
  & $h=4$
  & $h=5$
  & $h=6$
  & $h=7$
  & $h=8$
  & $h=9$
  & $h=10$
\\ \cmidrule(r){2-11}
$^\dagger$Linear Extrapolation
  & $0.65_{(0.06)}$
  & $0.79_{(0.05)}$
  & $0.93_{(0.05)}$
  & $1.08_{(0.05)}$
  & $1.24_{(0.05)}$
  & $1.39_{(0.05)}$
  & $1.55_{(0.05)}$
  & $1.70_{(0.05)}$
  & $1.86_{(0.05)}$
  & $2.01_{(0.05)}$
\\
$^\dagger$GP(SE$+$PER$+$WN)
  & $0.53_{(0.04)}$
  & $0.60_{(0.03)}$
  & $0.66_{(0.03)}$
  & $0.71_{(0.03)}$
  & $0.75_{(0.02)}$
  & $0.79_{(0.02)}$
  & $0.82_{(0.02)}$
  & $0.85_{(0.02)}$
  & $0.87_{(0.02)}$
  & $0.89_{(0.02)}$
\\
$^\dagger$GP(SE$\times$PER$+$WN)
  & $0.50_{(0.04)}$
  & $0.57_{(0.03)}$
  & $0.62_{(0.03)}$
  & $0.67_{(0.02)}$
  & $0.71_{(0.02)}$
  & $0.74_{(0.02)}$
  & $0.78_{(0.02)}$
  & $0.81_{(0.02)}$
  & $0.84_{(0.02)}$
  & $0.86_{(0.02)}$
\\
$^\dagger$Facebook Prophet 
  & $0.83_{(0.04)}$
  & $0.84_{(0.03)}$
  & $0.85_{(0.02)}$
  & $0.85_{(0.02)}$
  & $0.85_{(0.02)}$
  & $0.86_{(0.02)}$
  & $0.86_{(0.02)}$
  & $0.87_{(0.02)}$
  & $0.87_{(0.01)}$
  & $0.87_{(0.01)}$
\\
$^\dagger$Seasonal ARIMA 
  & $0.64_{(0.04)}$
  & $0.76_{(0.03)}$
  & $0.84_{(0.03)}$
  & $0.92_{(0.03)}$
  & $0.98_{(0.03)}$
  & $1.04_{(0.02)}$
  & $1.08_{(0.02)}$
  & $1.13_{(0.02)}$
  & $1.16_{(0.02)}$
  & $1.19_{(0.02)}$
\\
$^\dagger$TRCRP Mixture
  & $0.54_{(0.04)}$
  & $0.58_{(0.03)}$
  & $0.62_{(0.02)}$
  & $0.67_{(0.02)}$
  & $0.71_{(0.02)}$
  & $0.76_{(0.02)}$
  & $0.80_{(0.02)}$
  & $0.83_{(0.02)}$
  & $0.86_{(0.02)}$
  & $0.89_{(0.02)}$
\\
$^\ddagger$HDP-HSMM 
  & $0.69_{(0.05)}$
  & $0.72_{(0.04)}$
  & $0.76_{(0.03)}$
  & $0.79_{(0.03)}$
  & $0.82_{(0.02)}$
  & $0.84_{(0.02)}$
  & $0.86_{(0.02)}$
  & $0.88_{(0.02)}$
  & $0.89_{(0.02)}$
  & $0.90_{(0.02)}$
\\
$^\star$Multi-output GP 
  & $0.70_{(0.04)}$
  & $0.77_{(0.03)}$
  & $0.84_{(0.03)}$
  & $0.88_{(0.03)}$
  & $0.91_{(0.02)}$
  & $0.93_{(0.02)}$
  & $0.95_{(0.02)}$
  & $0.97_{(0.02)}$
  & $0.99_{(0.02)}$
  & $1.01_{(0.02)}$
\\
$^\star$TRCRP Mixture
  & $\mathbf{0.46}_{(0.03)}$
  & $\mathbf{0.49}_{(0.02)}$
  & $\mathbf{0.51}_{(0.02)}$
  & $\mathbf{0.53}_{(0.02)}$
  & $\mathbf{0.56}_{(0.02)}$
  & $\mathbf{0.58}_{(0.02)}$
  & $\mathbf{0.59}_{(0.01)}$
  & $\mathbf{0.61}_{(0.01)}$
  & $\mathbf{0.62}_{(0.01)}$
  & $\mathbf{0.64}_{(0.01)}$
\\ \bottomrule
\end{tabular*}
\scriptsize Modeled time series:
  \quad $^\dagger$flu
  \quad $^\ddagger$flu+weather
  \quad $^\star$flu+weather+tweets
\end{subtable}
\captionsetup{skip=5pt}
\caption{Quantitative evaluation of forecasting performance on the 2015 flu
season.
The table shows mean prediction errors and (one standard error) of the flu rate,
for various forecast horizons averaged over US Regions 1--10.
Available covariate time series include minimum temperature and Twitter messages
about the flu (not shown, see
Figure~\ref{subfig:multivariate-generative-model-flu}).
Predictive improvement of the multivariate TRCRP mixture over baselines is
especially apparent at longer horizons.
The top two panels show sample forecasts in US Region 6 for week 2014.51
(pre-peak) and week 2015.10 (post-peak).
The TRCRP mixture accurately forecasts seasonal dynamics in both cases, whereas
baseline methods produce inaccurate forecasts and/or miscalibrated
uncertainties.}
\label{fig:forecasting-baselines}
\end{figure*}

In this section, we give the full model likelihood and briefly describe MCMC
algorithms for inference in the hierarchical TRCRP mixture
\eqref{eq:crp-over-time-series}.
Since the model learns $M = \max(\bc^{1:N})$ separate TRCRP mixtures (one for
each time series cluster) we superscript latent variables of
Figure~\ref{subfig:multivariate-generative-model-math} by $m = 1,\dots,M$.
Namely, $\alpha^m$ is the CRP concentration, and $\z^m_{1:T}$ the latent regime
vector, shared by all time series in cluster $m$.
Further, let $K_m = \max(\z^m_{1:T})$ denote the number of unique regimes in
$\z^m_{1:T}$. Given window size $p$ and initial observations $\set{\x_{-p+1:0}^n :
n=1,\dots,N}$, we have:
\begin{flalign}
& \begin{array}{@{}l} P\Big(
  \alpha_0, \bc^{1:N},\; \alpha^{1:M},
  \lambda_G^{1:N},
  \lambda_F^{1:N},
  \set{\theta^n_j: 1 {\le}j{\le}K_{c^n}}_{n=1}^N, \\
  \qquad \z^{1:M}_{1:T},
  \x_{1:T}^{1:N} \; ;\;
  \x_{-p+1:0}^{1:N}, p \Big)
  \end{array} \notag \\
&= \Gamma(\alpha_0;1,1)\crp({\bc^{1:N} \mid \alpha_0}) \notag \\
& \;\; \left(\prod_{n=1}^N H^n_{G}(\lambda^n_G) \right)
  \left(\prod_{n=1}^NH^n_{F}(\lambda^n_F) \right)
  \left(\prod_{n=1}^N
    \prod_{j=1}^{K_{c^n}} \pi^n_{\Theta}(\theta_j^n)
  \right) \notag \\
& \;\; \prod_{m=1}^M \Bigg(
  \Gamma(\alpha^m;1,1) \prod_{t=1}^T
    \bigg[ b^m_t \crp(z^m_{t} \mid \z^m_{1:t-1}, \alpha^m)
    \notag \\
& \;\;\; \prod_{n|c_n=m} G(\x^n_{t-p:t-1}; D^n_{tz^m_t}, \lambda_G^n)
    F(x^n_t \mid \theta^n_{z^m_t}) \bigg] \Bigg)
    \label{eq:full-model-likelihood},
\end{flalign}
where $b_t^m$ normalizes the term between the square brackets, summed over
$z'^m_t=1,\dots,\max{(\z^m_{1:t-1})}+1$. Eq~\eqref{eq:full-model-likelihood}
defines the unnormalized posterior distribution of all latent variables given
the data. Appendix~\ref{appx:mcmc-methods} contains detailed algorithms for
posterior inference.
Briefly, temporal regime assignments $(z^m_{t}{|}\z^m_{1:T\backslash t}
\mathcomment{\dots})$ are sampled using a variant of Algorithm 3 from
\citep{neal2000}, taking care to handle the temporal-coupling term $b_t^m$ which
is not found in traditional DPM samplers.
We also outline an alternative particle-learning scheme \citep{carvalho2010} to
sample $(\z^m_{1:T}{|}\dots)$ jointly as a block.
Time series cluster assignments $(c^n{|}\bc^{1:N\backslash n},\dots)$ are
transitioned by proposing to move $\x^n$ to either an existing or a new cluster,
and computing the appropriate MH acceptance ratio for each case.
Model hyperparameters are sampled using an empirical Bayes approach
\citep{robbins1964} and the ``griddy Gibbs'' \citep{ritter1991} sampler.

\subsection{Making predictive inferences}
\label{subsec:making-predictive-inferences}

Given a collection of approximate posterior samples $\set{\hat\xi^1, \dots,
\hat\xi^S}$ of all latent variables produced by $S$ independent runs of MCMC, we
can draw a variety of predictive inferences about the time series $\x^{1:N}$
which form the basis of the applications in Section~\ref{sec:applications}.

\textbf{Forecasting} For out-of-sample time points, a forecast over an $h$ step
horizon $T < t < T + h$ is generated by ancestral sampling: first draw a chain
$\tilde{s} \sim \mathrm{Uniform}[1\dots S]$, then simulate step 5 of
Figure~\ref{subfig:multivariate-generative-model-math} using the latent
variables in chain $\xi^{\tilde{s}}$ for $t=T,\dots,T+h$.

\textbf{Clustering} For a pair of time series $(\x^i, \x^k)$, the posterior
probability that they are dependent is the fraction of samples in
which they are in the same cluster:
\begin{align}
\mathbb{P}\left[c^i = c^k \Big\vert \x^{1:N} \right] \approx
\frac{1}{S} \sum_{s=1}^S \mathbb{I}\left[\hat{c}^{i,s} = \hat{c}^{k,s}\right].
\label{eq:dependence-probability}
\end{align}
\textbf{Imputation} Posterior inference yields samples of each temporal regime
$\hat{z}^{\cdot,s}_{t}$ for all in-sample time points $1 \le t \le T$; the
posterior distribution of a missing value is:
\begin{align}
\mathbb{P}\left[x^n_t \in B \Big\vert \x^{1:N} \setminus \set{x^n_t} \right]
  \approx \frac{1}{S}
    \sum_{s=1}^S F(B \mid \hat\theta^{n,s}_{\hat{z}^{\hat{c}^{n,s}}_t}).
\label{eq:imputation}
\end{align}

\section{Applications}
\label{sec:applications}

In this section, we apply the TRCRP mixture to clustering hundreds of time
series using macroeconomic data from the Gapminder Foundation, as well as
imputation and forecasting tasks on seasonal flu data from the US Center for
Disease Control and Prevention (CDC).
We describe the setup in the text below, with further commentary given in
Figures~\ref{fig:cluster-datasets-gdp}, \ref{fig:imputation-baselines}, and
\ref{fig:forecasting-baselines}.
Experimental methods are detailed in Appendix~\ref{appx:experimental-methods}%
\footnote{An implementation of the hierarchical TRCRP mixture is available at
\url{https://github.com/probcomp/trcrpm}.}.

We first applied the TRCRP mixture with hierarchical prior to cluster countries
in the Gapminder dataset, which contains dozens of macroeconomic time series for
170 countries spanning 50 years.
Because fluctuations due to events such as natural disasters, financial crises,
or healthcare epidemics are poorly described by parametric or hand-designed
causal models, a key objective is to automatically discover the number and
kinds of patterns underlying the temporal structure.
Figure~\ref{fig:cluster-datasets-gdp} shows the outcome of structure discovery
in GDP time series using the model with $p=5$ years.
Several common-sense, qualitatively distinct clusters are detected.
Note that countries within each cluster share similar political, economic,
and/or geographic characteristics; see caption for additional details.
Appendix~\ref{appx:expanded-clustering} gives an expanded set of clusterings
showing changepoint detection in cell phone subscription time series, and
compares to a baseline using k-medoids clustering.

Predicting flu rates is a fundamental objective in public health policy.
The CDC has an extensive dataset of flu rates and associated time series such as
weather and vaccinations.
Measurements are taken weekly from January 1998 to June 2015.
Figure~\ref{subfig:multivariate-generative-model-flu} shows the
influenza-like-illness rate (ILI, or flu), tweets, and minimum
temperature time series in US Region 4, as well as six temporal
regimes detected by one posterior sample of the TRCRP mixture model
($p=10$ weeks).
We first investigated the performance of the proposed model on a multivariate
imputation task. Windows of length 10 were dropped at a rate of 5\% from flu
series in US Regions 1-10.
The top panel of Figure~\ref{subfig:imputation-baselines-imputations} shows flu
time series for US Regions 2, 4, 7, and 9, as well joint imputations (and two
standard deviations) obtained from the TRCRP mixture using
\eqref{eq:imputation}.
Quantitative comparisons of imputation accuracy to baselines are given in
Table~\ref{subfig:imputation-baselines-data}.
In this application, the TRCRP mixture achieves comparable accuracy to the
widely used Amelia II \citep{honaker2011} baseline, although neither method is
uniformly more accurate.
A sensitivity analysis showing imputation performance with varying $p$ is
given in Appendix~\ref{appx:p-sensitivity-imputation}.

To quantitatively investigate the forecasting abilities of the model, we next
held out the 2015 season for 10 US regions and generated forecasts on a rolling
basis.
Namely, for each week $t=2014.40,\dots,2015.20$ we forecast $\x^{\rm
flu}_{t:t+h}$ given $\x^{\rm flu}_{1:t-2}$ and all available covariate data up
to time $t$, with horizon $h=10$.
A key challenge is that when forecasting $\x^{\rm flu}_{t:t+h}$, the most recent
flu measurement is two weeks old $x^{\rm flu}_{t-2}$.
Moreover, covariate time series are themselves sparsely observed in the training
data (for instance, all Twitter data is missing before June 2013, top panel of
Figure~\ref{subfig:multivariate-generative-model-flu}).
Figure~\ref{fig:forecasting-baselines} shows the forecasting accuracy from
several widely-used, domain-general baselines that do not require detailed
custom modeling for obtaining forecasts, and that have varying ability to make
use of covariate data (weather and tweet signals).
The TRCRP mixture consistently produces the most accurate forecasts for all
horizons (last row).
Methods such as seasonal ARIMA \citep{hyndman2008} can handle covariate data in
principle, but cannot handle missing covariates in the training set or over the
course of the forecast horizon.
Both Facebook Prophet \citep{taylor2017} and ARIMA incorrectly forecast the peak
behavior (Figure~\ref{fig:forecasting-baselines}, top row), and are biased in
the post-peak regime (bottom row).
The HDP-HSMM \citep{johnson2013} also accounts for weather data, but fails to
detect flu peaks.
The univariate TRCRP (only modeling the flu) performs similarly to
periodic Gaussian processes, although the latter gives wider posterior error
bars, even in the relatively noiseless post-peak regime.
The multi-output GP \citep{alvarez2009} uses both weather and tweet covariates,
but they do not result in an improvement in predictive accuracy over univariate
methods.

\section{Discussion}
\label{sec:discussion}

This paper has presented the temporally-reweighted CRP mixture, a domain-general
nonparametric Bayesian method for multivariate time series.
Experiments show strong quantitative and qualitative results on multiple
real-world multivariate data analysis tasks, using little to no custom modeling.
For certain application domains, however, predictive performance may improve by
extending the model to include custom knowledge such as time-varying
functionals.
Further avenues for research include guidelines for selecting the window size;
greater empirical validation; a stick breaking representation; improving
inference scalability; and establishing theoretical conditions for posterior
consistency.
Also, it could be fruitful to integrate this method into a probabilistic
programming platform \citep{saad2016}, such as BayesDB.
This integration would make it easy to query mutual information between time
series \cite{saad2017}, identify data that is unlikely under the model, and make
the method accessible to a broader audience.

\section*{Acknowledgments}

This research was supported by DARPA PPAML program, contract number
FA8750-14-2-0004. The authors wish to thank Max Orhai from Galois, Inc. for
assembling the CDC flu dataset.

\bibliographystyle{abbrvnat}
\bibliography{main}

\clearpage

\begin{appendices}

\section{Data-dependent parameters for Student-T reweighting function}
\label{appx:data-dependent-reweighting}

Following \eqref{eq:reweighting-function-hypers}, the reweighting function $G$ is a
product of $p$ Student-T distributions whose location, scale and degrees of
freedom are data-dependent \citep{murphy2007}:
\begin{align*}
&G(\x_{t-p:t-1}; D_{tk}, \lambda_G)                             \\
& = \prod_{i=1}^{p} G_i(x_{t-i} ; D_{tki}, \lambda_{Gi})        \\
& = \prod_{i=1}^{p} \mathrm{T}_{2a_{tki}} \left(
    x_{t-i} ; m_{tki}, b_{tki}\frac{V_{tki}+1}{a_{tki}}
    \right) \ttag \label{eq:reweighting-function-hypers-restate}   \\
\lambda_{Gi}  &= (m_{i0}, V_{i0}, a_{i0}, b_{i0})               \\
D_{tki}       &= \set{x_{t'-i} : z_{t'} = k, 1 \le t' < t}      \\
n_{tki}       &= \lvert D_{tki} \rvert \\
\bar{x}_{tki} &= \frac{1}{n_{tki}}\textstyle\sum_{t'\in D_{tki}} {x_{t'-i}}
              \ttag \label{eq:reweighting-function-hypers-updates}         \\
V_{tki}       &= 1/(V_{i0}^{-1} + n_{tki})                              \\
m_{tki}       &= V_{tki}(V_{i0}^{-1}m_{i0} + n_{tki}\bar{x}_{tki})      \\
a_{tki}       &= a_{i0} + n_{tki}/2                                     \\
b_{tki}       &= b_{k0} + \frac{1}{2}\left(
                m_{i0}^2V_{i0}^{-1} + \textstyle\sum_{t'}x^2_{t'-i}
                -m^2_{itk}V^{-1}_{tki}
              \right).
\end{align*}

\section{Markov chain Monte Carlo methods for posterior inference}
\label{appx:mcmc-methods}

Here, we provide the details of the MCMC method for posterior simulation from
the nonparametric mixture model developed in
Section~\ref{sec:temporally-reweighted-crp}. As discussed in the main text,
conjugacy of $F$ and $\pi_\Theta$ in \eqref{eq:normal-inverse-gamma} means we
can analytically marginalize parameters $\set{\theta^n_k}$ when defining the
generative process of the TRCRP mixture. The model in
Figure~\ref{subfig:multivariate-generative-model-math} therefore becomes:
\begin{flalign*}
& \alpha \sim \textrm{Gamma(1,1)} \ttag \label{eq:generative-process-appx}
  \span \\
& \lambda_G^n \sim H_G^n
  & n=1,2,\dots,N \\
& \lambda_F^n \sim H_F^n
  & n=1,2,\dots,N \\
& \x^n_{-p+1:0} \defeq (x^n_{-p+1}, \dots, x^n_{0})
  & n=1,2,\dots,N \\
& {\Pr}\left[
      z_t = k \mid \z_{1:t-1}, \x^{1:N}_{t-p:t-1}, \alpha, \lambda^{1:N}_G
    \right]
  & t=1,2,\dots,T \\
& \quad \propto {\crp}(k|\alpha, \z_{1:t-1})
    \textstyle\prod_{n=1}^N G(\x^n_{t-p:t-1}; D^n_{tk}, \lambda_G^n)
  \span \\
& \quad \mathcomment{where }
    D^n_{tk} \defeq \set{\x^n_{t'-p:t'-1} \mid z_{t'} = k, 1 \le {t'} < t}
  \span \\
& \quad  \mathcomment{and } k=1,\dots, \max{(\z_{1:t-1})}+1
  \span \\
& x^n_t \big\vert \set{z_t = k, \x^{1:N}_{1:t-1}} \sim
    \textstyle\int_{\theta} F(\cdot \vert \theta)
      \pi_{\Theta}(\theta | D'^n_{tk}, \lambda^n_F)\mathrm{d}\theta
  \span \\
& \quad \mathcomment{where }
    D'^n_{tk} \defeq \set{\x^n_{t'} \mid z_{t'} = k, 1 \le {t'} < t}.
  \span \\
& & n=1,2,\dots,N
\end{flalign*}
The integration of $F$ against $\pi_\Theta(\theta|D'^n_{z_t})$ in the right
hand-side of the final line evaluates to a Student-T distribution as in
\eqref{eq:reweighting-function-hypers-restate}, whose updates given
$D'^n_{tz_t}$ and $\lambda^n_F$ are identical to those in
\eqref{eq:reweighting-function-hypers-updates} with $i=0$.

\textbf{Inference on temporal regime assignments} $(z_{t}{|}\z_{1:T\backslash
t}, \mathcomment{\dots})$. We first describe how to transition $\z_{1:T}$,
assuming the collapsed version of the TRCRP
\eqref{eq:generative-process-appx} with $N$ time series.
Note that since the hierarchical prior \eqref{eq:crp-over-time-series} for
structure learning results in $M = \max(\bc^{1:N})$ independent TRCRP mixtures
(conditioned on the assignment vector), it suffices to describe inference on
$\z_{1:T}$ in one of the mixtures (which keeps notation significantly simpler).
Given observations $\x^{1:N}_{-p+1:T}$, the joint likelihood of model
\eqref{eq:generative-process-appx} is:
\begin{flalign*}
& P\left(
  \alpha, \lambda_G^{1:N}, \lambda_F^{1:N}, \z_{1:T},
  \x_{1:T}^{1:N} \; ; \; \x_{-p+1:0}^{1:N}, p \right) & \\
&= \Gamma(\alpha;1,1) \left(\prod_{n=1}^N H^n_{G}(\lambda^n_G) \right)
  \left(\prod_{n=1}^NH^n_{F}(\lambda^n_F) \right) & \\
&\qquad \prod_{t=1}^T  \bigg[ b_t \crp(z_{t} \mid \z_{1:t-1}, \alpha) \notag & \\
&\qquad\; \prod_{n=1}^N
    G(\x^n_{t-p:t-1}; D^n_{tz_t}, \lambda^n_G)
    F(x^n_t \mid D'^n_{tz_t}, \lambda^n_F) \bigg]
    & \ttag \label{eq:full-model-likelihood-appx}
\end{flalign*}
The normalizer at time $t$ is given by:
\begin{align}
& b_t(\x^{1:N}_{1:t-1}, \z_{1:t-1}) \label{eq:normalizer} \\
&\; = \left(\sum_{k=1}^{K_t} {\crp}(k|\alpha, \z_{1:t-1})
    \prod_{n=1}^N G(\x^n_{t-p:t-1}; D^n_{tk}, \lambda_G^n) \right)^{-1}, \notag
\end{align}
where $K_t = \max(\z_{1:t-1}) + 1$. Note that the normalizer
$b_t(\x^{1:N}_{1:t-1}, \z_{1:t-1})$ ensures the reweighted cluster probabilities
sum to one.
It will also be convenient to define the predictive density $q_t$ at time $t$ of
data $\x^{1:N}_t$, which sums out all possible values of $z_t$:
\begin{align*}
& q_t(\x^{1:N}_{1:t}, \z_{1:t-1}) \ttag \label{eq:normalizer-predictive} \\
&= b_t(\x^{1:N}_{1:t-1}, \z_{1:t-1}) \left(
  \sum_{k=1}^{K_t} {\crp}(k|\alpha, \z_{1:t-1})
\right. \\
&\qquad \left. \prod_{n=1}^N
  G(\x^n_{t-p:t-1}; D^n_{tk}, \lambda_G^n)
  F(x^n_{t} \mid D'^n_{tk}, \lambda_F^n)
\right).
\end{align*}

Let the current state of the Markov chain be $(\alpha, \lambda_G^{1:N},
\lambda_F^{1:N}, \z_{1:T})$. We present two algorithms for sampling the latent
regimes assignments.
Algorithm~\ref{alogri:mh} is a single-site Metropolis-Hastings
procedure that targets $(z_t {\mid} \z_{1:T\backslash{t}}, \dots)$ at each step,
where we assume that all data in $\x^{1:N}_{1:T}$ are fully observed.
Algorithm~\ref{algori:smc} is an SMC scheme to block sample
$(\z_{1:T}\vert\dots)$ using particle learning \citep{carvalho2010}. Arbitrary
observations may be missing, as they are imputed over the course of
inference.

\begin{figure*}

\begin{framed}
\refstepcounter{algori}\label{alogri:mh}%
\textit{Algorithm \ref{alogri:mh}: single-site Metropolis-Hastings}. This
algorithm proposes $(z_t {\mid} \z_{1:T\backslash{t}}, \dots)$ at each step,
assuming fully observed data $\x^{1:N}_{1:T}$. Repeat for $t=1,2,\dots,T$:
\begin{enumerate}[leftmargin=*]
\item Propose $z'_t$ from the multinomial distribution:
\begin{flalign}
&\Pr[z'_t = k \mid \z_{1:T\backslash{t}}, \x^{1:N}, \alpha]
\propto {\crp}(k|\alpha, \z_{1:T\backslash{t}})
\prod_{n=1}^N
  G(\x^n_{t-p:t-1}; D^n_{Tk}\backslash\set{x^n_t}, \lambda_G^n)
F(x^n_{t} \mid D'^n_{Tk}\backslash\set{x^n_t}, \lambda_F^n),
\label{eq:computing-zt-proposal} \\
&\qquad \mathcomment{for }
  k \in \mathrm{unique}(\z_{1:T\backslash{t}}) \cup
  \set{\max(\z_{1:T\backslash{t}})+1}. \notag
\end{flalign}

\item Compute the MH acceptance ratio $r(z_t \to z'_t)$, using $b_t$
defined in \eqref{eq:normalizer}:
\begin{flalign}
r(z_t \to z'_t) = \frac{
  \prod_{t'>t} b_{t'}(\z_{1:t'-1\backslash{t}} \cup z'_t, \x^{1:N}_{1:t'-1})}{
  \prod_{t'>t} b_{t'}(\z_{1:t'}, \x^{1:N}_{1:t'-1})}{
}. \label{eq:computing-zt-acceptance}
\end{flalign}

\item Set $z_t \gets z'_t$ with probability $\min(1,r)$, otherwise leave $z_t$
unchanged.
\end{enumerate}
\end{framed}

\begin{framed}
\refstepcounter{algori}\label{algori:smc}%
\textit{Algorithm \ref{algori:smc}: block sampling with particle-learning}. This
algorithm block samples $\z_{1:T}$ without any assumptions on missingness of
observations. Let $o^n_t$ be the ``observation indicator'' so that $o^n_t = 1$
if $x^n_t$ is observed, and 0 if it missing ($n = 1,2,\dots,N$ and
$t=1,2,\dots,T$). Let $J > 0$ be the number of particles. Since we will be
simulating missing values over the course of inference, we superscript all data
with $j$ to indicate the inclusion of any imputed values by particle $j$.
\begin{enumerate}[label*=\arabic*.,leftmargin=*]
\item Set $w^j \gets 1$ for $j=1,2,\dots, J$

\item Repeat for $t=1,2,\dots,T$

\begin{enumerate}[label*=\arabic*.,leftmargin=*]
\item Repeat for $j=1,2,\dots,J$

  \begin{enumerate}[label*=\arabic*.,leftmargin=*]

  \item Sample $z^j_t$ from the multinomial distribution:
  \begin{align}
  \Pr[z^j_t = k \mid \z^j_{1:t-1}, \x^{1:N,j}, \alpha]
  & \propto {\crp}(k|\alpha, \z^j_{1:t-1})
  \prod_{n=1}^N G(\x^{n,j}_{t-p:t-1}; D^{n,j}_{tk}, \lambda_G^n) \notag \\
  &\qquad \prod_{n=1}^N\left(
    F(x^n_{t} \mid D'^{n,j}_{tk}, \lambda_F^n)
  \right)^{o^n_t},
  \label{eq:smc-sample-cluster} \\
  \mathcomment{for } k &= 1, 2, \dots, \max(\z^j_{1:t-1}) + 1. \notag
  \end{align}

  \item Update particle weight using predictive density $q_t$ defined in
  \eqref{eq:normalizer-predictive}:
  \begin{align}
  w^j \gets w^j q_t\left(
    \x^{1:N,j}_{1:t-1} \cup \set{x^n_t \mid o^n_t = 1}, \z^j_{1:t-1}
  \right).
  \label{eq:update-particle-weight}
  \end{align}

  \item For each $n$ such that $o^n_t = 0$, simulate a value
  $x^{n,j}_t \sim F(\cdot \mid D'^n_{tz^j_t}, \lambda_F^n)$.
  \end{enumerate}

\item If resampling criterion met, then:

  \begin{enumerate}[label*=\arabic*.,leftmargin=*]
  \item Resample $(\z^j_{1:t}, \x^{1:N,j}_{1:t})$ proportionally to $w^j,
    \, j=1,2,\dots,J$.
  \item Renormalize weights $w^j \gets w^j/\sum_{j'}w^{j'},
    \, j=1,2,\dots,J$.

  \end{enumerate}
\end{enumerate}

\item Resample $j \sim \mathrm{Categorical}(w^1,\dots,w^J)$ and return
$(\z^j_{1:T}, \x^{1:N,j}_{1:T})$.
\end{enumerate}
\end{framed}
\end{figure*}

\FloatBarrier

It is worth discussing the computational trade-offs between MH
Algorithm~\ref{alogri:mh} and SMC Algorithm~\ref{algori:smc}. In step 1 of
Algorithm~\ref{alogri:mh}, \eqref{eq:computing-zt-proposal} is recomputed
$K=O(\max(\z_{1:T}))$ times. Each assessment requires $O(Np)$ computations,
where the factor of $N$ is the product over the time series, and the factor of
$p$ is the cost of assessing $G$ per \eqref{eq:reweighting-function-hypers}. In
step 2, computing the terms $b_{t'}$ in the acceptance ratio
\eqref{eq:computing-zt-acceptance} requires revisiting $O(T)$ data points.
Therefore a single iteration requires $O(TKNp)$ computations, so that the cost
of a full sweep over all $T$ time points is $O(T^2KNp)$.
Note that it is not necessary to sum over $K_t$ in
\eqref{eq:normalizer} when computing the $b_{t'}$ terms in
\eqref{eq:computing-zt-acceptance}, since the data in at most two clusters will
change when proposing $z_t$ to $z_{t'}$. The sufficient statistics can be
updated in constant time using a simple dynamic programming approach.

In practice, we consider several computational approximations that simplify the
scaling properties of the single-site MH Algorithm~\ref{alogri:mh}. For missing
data, rather than evaluate the full model likelihood
\eqref{eq:full-model-likelihood-appx} on imputed data for each $t = 1,
\dots, T$, we instead adopt a ``data-dependent'' prior, similar to the strategy
described by \citep{dunson2007} in the context of Bayesian density regression.
Namely, letting $o^n_t$ be the indicator for having observed $x^n_t$, we let the
reweighting function $G$ consider only those data points that have actually been
observed. Therefore, \eqref{eq:reweighting-function-hypers} becomes:
\begin{align}
G(\x_{t-p:t-1}; D_{tk}, \lambda_G)
= \prod_{i=1}^{p} (G_i(x_{t-i} ; D_{tki}, \lambda_{Gi}))^{o^n_{t-i}}.
\label{eq:missing-data-heuristic}
\end{align}
Second, note that the MH proposal \eqref{eq:computing-zt-proposal} is very
similar to the Gibbs proposal from Algorithm 3 of \citep{neal2000}, except we
must account for the temporal coupling so that the transition is guaranteed to
leave \eqref{eq:full-model-likelihood-appx} invariant. Empirical evidence
suggest that, when using the proposal \eqref{eq:computing-zt-proposal},
acceptance ratios center around one. This observation suggests a good
initialization strategy for the Markov chain (prior to running the full MH
algorithm): run several rounds of step 1 always accepting the proposal $z_t \to
z'_t$ without computing \eqref{eq:computing-zt-acceptance}, which eliminates the
additional $O(T)$ factor.

Unlike the MH Algorithm~\ref{alogri:mh}, the SMC algorithm
\eqref{algori:smc} with requires $O(KNp)$ to assess \eqref{eq:smc-sample-cluster}
in step 2.1.1; the total cost of a complete pass through all $T$ data points
(step 2) and all $J$ particles (step 2.1) is therefore $O(JTKNp)$. Note that in
SMC, the normalizers $b_t$ need not to be retroactively computed, which is the
key overhead of MH.
In addition to its linear scaling in $T$, SMC is able to (i) more tractably
handle missing data, and (ii) use a posterior particle filter by sampling from
the conditionally optimal proposal distribution in step 2.1.1, resulting in
significantly lower variance of the weights \citep{carvalho2010}.

\textbf{Inference on time series cluster assignments} %
$(c^n{|}\bc^{1:N\backslash n},\dots)$. %
This section describes an MCMC algorithm for sampling the time series cluster
assignments when using the hierarchical CRP structure prior
\eqref{eq:crp-over-time-series}.
For notational simplicity, let $B \subseteq [N]$ and define:
\begin{align*}
& L^m(\z_{1:T}, \x^{B}_{1:T}) = \prod_{t=1}^T \bigg[
  b_t \crp(z_{t} \mid \z_{1:t-1}, \alpha^m) \\
&\quad\; \prod_{n=1}^N
    G(\x^n_{t-p:t-1}; D^n_{tz_t}, \lambda^n)
    F(x^n_t \mid D'^n_{tz_t}, \lambda^n_F) \bigg].
    \ttag \label{eq:partial-model-likelihood-appx}
\end{align*}
The term $L^m$ is a short-hand for the product from $t=1$ to $T$ in the full
model likelihood \eqref{eq:full-model-likelihood-appx} for a single TRCRP
mixture, with latent sequence $\z_{1:T}$, data $\x^B_{1:T}$, and CRP
concentration $\alpha^m$. Second, let $A^m = \set{n \mid c^n = m}$ be the
indices of the time series currently assigned to cluster $m$.

\begin{framed}
\refstepcounter{algori}\label{algori:view}%
\textit{Algorithm \ref{algori:view}: Sampling time series cluster assignments}.
Let the current state of the Markov chain be
$( \alpha_0, \bc^{1:N}, \alpha^{1:M}, \lambda_G^{1:N}, \lambda_F^{1:N},
\z^{1:M}_{1:T})$
with observations $\x^{1:N}_{1:T}$. This algorithm resamples
$(c^n{|}\bc^{1:N\backslash n},\dots)$. Repeat for $n=1,2,\dots,N$:

\begin{enumerate}[label*=\arabic*.,leftmargin=*]

\item If $c^n$ is not a singleton cluster, i.e. $\lvert A^{c^n} \rvert > 1$,
then generate a proposal sequence by forward sampling $\z^{M+1}_{1:T}$ from
model prior \eqref{eq:generative-process-appx}, holding the data $\x^n_{1:T}$
fixed at the observed values.

\item If $c^n$ is a singleton, i.e. $\lvert A^{c^n} \rvert = 1$, then re-use
the current latent regime sequence by setting $\z^{M+1}_{1:T} = \z^{c^n}_{1:T}$

\item For $m \in \mathrm{unique}(\bc^{1:N\backslash{n}})$, compute
  \begin{align*}
  p^m = \begin{cases}
    \lvert A^{m} \rvert L^m\left(\z^m_{1:T}, \x^n_{1:T}\right)
     & \mathcomment{if } c^n \ne m, \\
    (\lvert A^{m} \rvert - 1) L^m\left(\z^m_{1:T}, \x^n_{1:T}\right)
     & \mathcomment{if } c^n = m.
  \end{cases}
  \end{align*}%

\item Compute the singleton proposal probability:
  \begin{align*}
  p^{M+1} = \alpha_0 L_{m+1}\left(\z^{M+1}_{1:T}, \x^n_{1:T}\right)
  \end{align*}

\item Sample $c' \sim \mathrm{Categorical}(\set{p^m})$.

\item Compute the MH acceptance ratio
  \begin{align*}
  &r(c^n \to {c'}) = \\
  &\;\, \left(\frac{
    L^{c'}(\z^{c'}_{1:T}, \x^{A^{c'}}_{1:T}\cup\x^{n}_{1:T})
    L^{c^n}(\z^{c^n}_{1:T}, \x^{A^{c^n}}_{1:T}\backslash\x^{n}_{1:T})
  }{
    L^{c'}(\z^{c'}_{1:T}, \x^{A^{c'}}_{1:T})
    L^{c^n}(\z^{c^n}_{1:T}, \x^{A^{c^n}}_{1:T})
  } \right) \\
  &\;\, \left( \frac{
    L^{c^n}(\z^{c^n}_{1:T}, \x^{n}_{1:T})
  }{
    L^{c'}(\z^{c'}_{1:T}, \x^{n}_{1:T})
  } \right). \ttag \label{eq:computing-cn-acceptance}
  \end{align*}

\item Set $c^n \gets c'$ with probability $\min(1,r)$, else leave $c^n$
unchanged.

\end{enumerate}
\end{framed}

By proposing the latent regime singleton from the (conditional) prior in Step 2
of Algorithm~\ref{algori:view}, transdimensional adjustments such as reversible
jump MCMC \cite{green1995} need not be considered. Second, when computing the MH
acceptance ratio \eqref{eq:computing-cn-acceptance} in step 6, it is not
necessary to recompute all the $L^m$ terms at each iteration. First, writing out
the full products \eqref{eq:partial-model-likelihood-appx} results in
cancellation of several terms in the numerator and denominator of
\eqref{eq:computing-cn-acceptance}. Second the $b^m_t$ terms that do not cancel
contain several duplicated components, which can be reused from one transition
to the other.

In practice, we find that a similar heuristic to the one described for
Algorithm~\ref{alogri:mh} provides good transitions in the state space, given
the similarities between Algorithm~\ref{algori:view} and the Gibbs Algorithm 8
from \citep{neal2000}.

\textbf{Inference on model hyperparameters} $(\alpha_0,\set{\alpha^m},
\set{\lambda^n_G}, \set{\lambda^n_F} \mid \dots)$. This section describes the
empirical Bayes approach \citep{robbins1964} for transitioning model
hyperparameters, using the ``griddy Gibbs'' approach from \citep{ritter1991}.
For each hyperparameter, we construct a grid of 30 data-dependent
logarithmically-spaced bins as follows:
\begin{align*}
\shortintertext{\textrm{\underline{Outer CRP concentration}}}
&\mathcomment{grid}(\alpha_0) &&= \mathcomment{logspace}(1/N, N) \\
\shortintertext{\textrm{\underline{TRCRP concentration}}}
&\mathcomment{grid}(\alpha^m) &&= \mathcomment{logspace}(1/T, T) \\
\shortintertext{\textrm{\underline{Normal-InverseGamma hyperparameters}}}
&\mathcomment{grid}(m^n_{0})  &&= \mathcomment{logspace}(
  \min(\x^n_{1:T})-5, \max(\x^n_{1:T})+5) \\
&\mathcomment{grid}(V^n_{0})  &&= \mathcomment{logspace}(1/T, T) \\
&\mathcomment{grid}(a^n_{0})  &&= \mathcomment{logspace}(
  \mathcomment{ssqdev}(\x^n_{1:T})/100, \mathcomment{ssqdev}(\x^n_{1:T})) \\
&\mathcomment{grid}(b^n_{0})  &&= \mathcomment{logspace}(1, T).
\end{align*}

Grids for the Normal-InverseGamma hyperparameters apply to both $\lambda_F$
$(n=1,2,\dots,N)$ and $\lambda_G$ (windows $i=1,2\dots,p$). We cycle through the
grid points of each hyperparameter, and assess the conditional likelihood at
each bin using \eqref{eq:full-model-likelihood}. We find that this method is
both computationally reasonable and finds good hyperparameter settings. However,
alternative approaches based on slice sampling offer a promising alternative to
achieve fully Bayesian inference over hyperparameters.

\section{Experimental Methods}
\label{appx:experimental-methods}

This section describes the quantitative experimental methods used for
forecasting, clustering, and imputation pipelines in
Section~\ref{sec:applications}. Access to experimental pipeline code is
available upon request.

\subsection{Flu forecasting}
\label{appx:experimental-methods-flu-forecasting}

The full CDC flu datasets used in this paper are available at
\url{https://github.com/GaloisInc/ppaml-cp7/tree/master/data}. Flu populations
were constructed from the following csv files: \textsf{USA-flu.csv},
\textsf{USA-tweets.csv}, and \mbox{\textsf{USA-weather.csv}}.
In each of US Regions 1 through 10, we held out data from weeks 2014.40 through
2015.20, and produced forecasts with a 10 week horizon on a rolling basis. Tweet
and minimum temperature covariates were used. More precisely, for a region
$r$ (such as US Region 10) a forecaster $F$ for week $t$ extending $h$ weeks into
the future is a function:
\begin{align}
F_{r,t,h} : \set{\x^{{\rm flu,}r}_{1:t-2}, \x^{{\rm cov},r}_{1:t}}
  \mapsto \set{\x^{{\rm flu},r}_{t:t+h}}.
\end{align}
The forecastors iterated over regions $r=1,2,\dots,10$, weeks
$t=2014.40,2014.41,\dots,2015.20$, and horizons $h=1,2,\dots,10$. Note that the
two week delay in the latest flu data is expressed by only having data up to
$t-2$ when forecasting at week $t$. Second, $\x^{\rm cov}$ contains arbitrary
missing values (see for example the tweets time series from
Figure~\ref{subfig:multivariate-generative-model-flu}). When forecasting,
covariate values are only available up to the current week $t$, not the entire
course of the forecast horizon. Nine forecasting methods were used in the paper,
shown in Figure~\ref{fig:forecasting-baselines}. Below are further details on
each forecaster:

\textbf{Constant}. This method returns a constant prediction based on the most
recently observed flu value $x^{\rm flu}_{t-2}$ over the entire course of
the horizon.

\textbf{Linear extrapolation}. This method fits a straight line through the
three most recently observed flu values, $\x^{\rm flu}_{t-4:t-2}$, and returns
predictions by extrapolating the line for $h$ weeks.

\textbf{GP (SE+PER+WN)}. This method is a Gaussian process whose covariance
kernel is a sum of squared exponential, periodic, and white noise components.
Hyperparameter inference was conducted using the open source implementation from
the Venture platform
\citep[\url{https://github.com/probcomp/Venturecxx}]{schaechtle2015}. %
MH sampling on data-dependent hyperparameter grids were run for a burn-in period
of 10000 iterations. Predictions were obtained by drawing 500 independent curves
from the posterior predictive distribution, evaluated jointly at the forecast
weeks.

\textbf{GP (SE${\times}$PER+WN)}. Identical to above, except to using a
covariance kernel with a product of squared exponential and periodic components,
plus white noise. The change in covariance kernel resulted in little
quantitative and qualitative differences.

\textbf{Facebook Prophet}. We used the open-source python implementation of
Facebook Prophet \citep[\url{https://facebook.github.io/prophet}]{taylor2017}.
We specified the data sampling rate as weekly. The method requires no additional
specification or tuning. The predictor returns point estimates, as well as upper
and lower confidence intervals, at the held-out weeks.

\textbf{Seasonal ARIMA}. We used the R implementation of seasonal ARIMA from the
\texttt{forecast} package %
\citep[\url{https://cran.r-project.org/web/packages/forecast}]{hyndman2008}. The
model is parameterized as ARIMA$(p,d,q)(P,D,Q)_m$, where $p$ is the non-seasonal
AR order, $d$ is the non-seasonal differencing, $q$ is the non-seasonal MA
order, $P$ is the seasonal AR order, $D$ is the seasonal differencing, $Q$ is
the seasonal MA order, and $m$ is the sampling frequency per period. For each of
the 10 flu seasons, we used \texttt{auto.arima} to perform model selection. We
manually specified the weekly sampling rate by setting $m=52$, and set $D=1$ to
specify 1 flu season per year. The program optimize all other parameters using
non-stepwise grid search, which is significantly slower to fit than stepwise
search, but is both more extensive and more appropriate for data with seasonal
behavior (according to the package documentation).  While \texttt{auto.arima}
can in principle support covariate data using the \texttt{xreg} parameter, we
were unable to successfully use \texttt{xreg} due to missing data in the matrix
of external regressors (tweets and weather) at the held-out weeks. The predictor
returns point estimates, as well as upper and lower confidence intervals, at the
held-out weeks.

\textbf{Multi-output GP} This method is a single-input (time) multiple-output
(flu, tweets, and weather data) Gaussian process. We used the the open source
MATLAB implementation of sparse convolved Gaussian process for multi-output
regression from the \texttt{multigp} package
\cite[\url{https://github.com/SheffieldML/multigp}]{alvarez2009}. We used the
following configuration options:
\begin{enumerate}[label=\roman*]
\setlength\itemsep{0ex}
\item \texttt{multigpOptions(\textquotesingle{ftc}\textquotesingle)};
\item \texttt{options.kernType{=}\textquotesingle{ggwhite}\textquotesingle};
\item \texttt{options.optimizer{=}\textquotesingle{scg}\textquotesingle};
\item \texttt{options.nlf{=}1},
\end{enumerate}
to specify (i) full estimation without running likelihood approximations; (ii) a
Gaussian-Gaussian kernel with white noise; (iii) scaled conjugate gradient
optimization; and (iv) one latent function.
Moreover, the \texttt{options.bias} and \texttt{options.scale} parameters were
initialized to their empirical values from the training set. Optimization was
run until convergence for all forecastors. This method is the only baseline
which can handle arbitrary patterns of missing data, thereby making use of the
weather and tweet signals when forecasting predictions at time $t$. However, the
absence of a periodic kernel in the convolved GP implementation made it
difficult to capture the seasonal dynamics. Predictions were obtained by
sampling 500 independent normal random variables from the posterior predictive
distribution evaluated at the forecast weeks.

\textbf{HDP-HSMM}. This method is the hierarchical Dirichlet process semi-Markov
model; experiments were run using the open-source python package \texttt{pyhsmm}
\citep[\url{https://github.com/mattjj/pyhsmm}]{johnson2013}. While the HDP-HSMM
cannot handle missing values in the training data, it can handle missing data
over the course of the prediction horizon. Therefore, flu and weather time
series were modeled jointly, leaving out the tweets. We used the
\texttt{WeakLimitHDPHSMM} model, with a Poisson duration distribution and
Gaussian observation distribution. Default configurations of all hyperparameters
of these distributions and the HDP-HSMM concentration were taken from examples
made available by the authors. MCMC inference with 1000 steps of burn-in was
used. Predictions were obtained by drawing 100 independent curves from the
posterior predictive evaluated at the forecast weeks.

\textbf{Univariate TRCRP mixture}. This method only considered the flu time
series using model \eqref{eq:crp-probability-reweighted}. We used a window size
of $p=10$ weeks, and $S=64$ parallel MCMC runs with a burn-in period of 5000
iterations. Predictions were obtained by drawing 500 independent curves from the
posterior predictive distribution evaluated at the forecast weeks.

\textbf{Multivariate TRCRP mixture}. This method considered flu, weather
and tweet time series using the model in
Figure~\ref{subfig:multivariate-generative-model-math}. We used a window size of
$p=10$ weeks, and $S=64$ parallel MCMC runs with a burn-in period of 5000
iterations. Missing covariate data was handled using the approximation given in
\eqref{eq:missing-data-heuristic}. Using the hierarchical structure prior
\eqref{eq:crp-over-time-series} resulted in little to no quantitative
difference. The three time series are dependent, which was reflected in their
posterior dependence probability
\eqref{eq:dependence-probability} being 1 across all 64 independent chains.
Predictions were obtained by sampling 500 independent curves from the posterior
predictive distribution evaluated at the forecast weeks. An open-source
implementation of the method used in this paper is at
\url{https://github.com:probcomp/trcrpm}.

\subsection{Flu imputation}
\label{appx:experimental-methods-flu-imputation}

We constructed a single population of 10 flu time series for US Regions 1
through 10. Missing data was dropped independently in each time series by
removing consecutive windows of length 10 at a rate of $5\%$. The full and
dropped datasets used for benchmarking are shown in
Figure~\ref{fig:observed-missing-data-all}. Below are further on details on each
of the five imputation methods:

\textbf{Mean imputation}. This method returns the per-series mean as the
imputed value for each data point.

\textbf{Linear interpolation}. This method constructs a straight line between
every pair of time points $t_1 < t_2$ which have at least one missing
observation between them. The interpolation method used was
\texttt{pandas.Series.interpolate} from the python \texttt{pandas} package at
\url{https://pandas.pydata.org}.

\textbf{Cubic interpolation}. The cubic interpolation routine used was
\texttt{scipy.interpolate.interp1d} from the python \texttt{scipy} package at
\url{https://scipy.org}.

\textbf{Amelia II}. This method uses the R package \texttt{amelia}
\citep[\url{http://cran.r-project.org/web/packages/Amelia}]{honaker2011} for
multiple imputation. We used 100 samples per missing data point. Imputation
errors were averaged over the multiple imputations.

\textbf{Multivariate TRCRP mixture}. A window of $p=10$ weeks was used, with
$S=64$ parallel MCMC runs and a burn-in period of 5000 iterations. 100
predictive samples from each of the chains were obtained using
\eqref{eq:imputation}, and imputation errors were averaged over the multiple
imputations. Joint imputations of Regions 1 through 10 are are shown
Figure~\ref{fig:observed-missing-data-all}.

\subsection{Sensitivity of imputation performance to the TRCRP mixture window
size}
\label{appx:p-sensitivity-imputation}

\begin{figure}[ht]
\centering
\includegraphics[width=\linewidth]{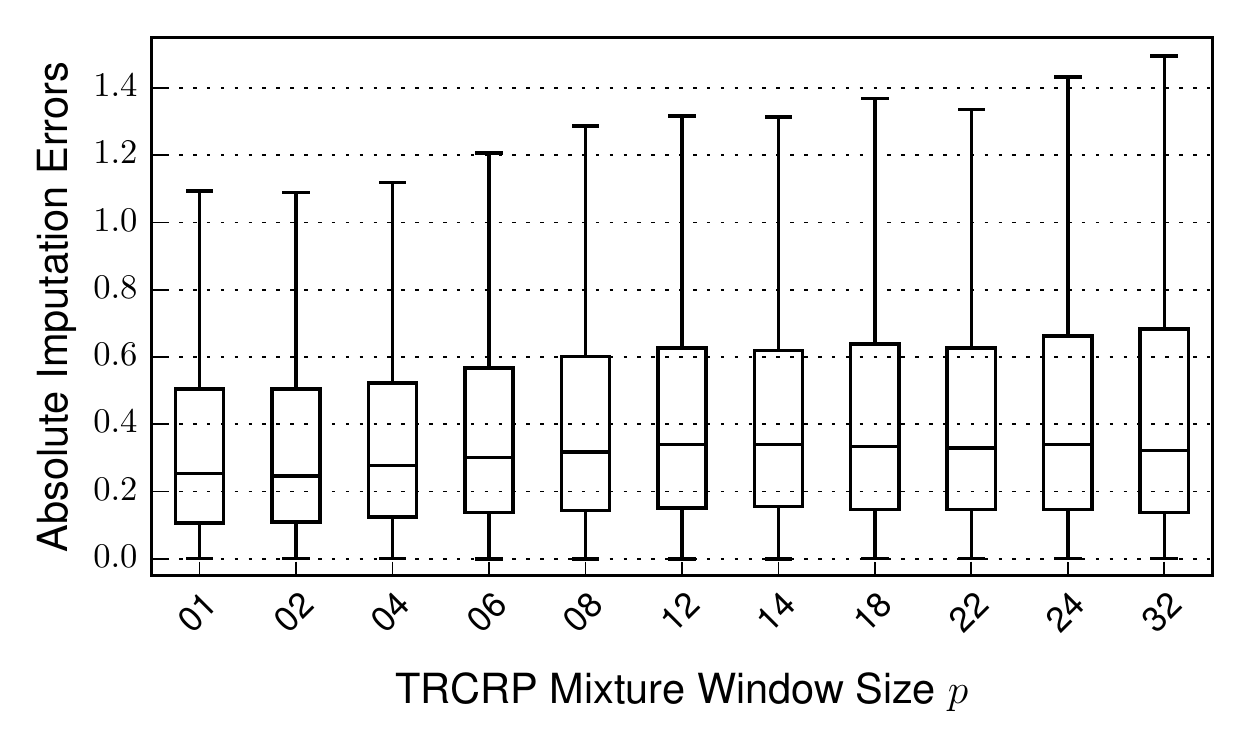}
\captionsetup{skip=0pt}
\caption{Sensitivity of imputation performance to TRCRP window size $p$.}
\label{fig:p-sensitivity-imputation}
\end{figure}

We further studied how imputation performance of the TRCRP mixture varied as we
changed the window size $p$. Figure~\ref{fig:p-sensitivity-imputation} shows the
outcome of this sensitivity analysis. In all cases, the sampler was run for a
burn-in of 5000 iterations with $S=16$ chains. While imputation is generally not
highly sensitive to $p$, median imputation values degrades slightly with
increasing $p$ and the variance of imputation errors increases. (At higher $p$,
the MCMC chains need a significantly higher number of iterations to mix well
than at lower $p$.)

\begin{figure*}[ht]
\begin{subfigure}{.5\linewidth}
\captionsetup{skip=0pt}
\includegraphics[width=\textwidth]{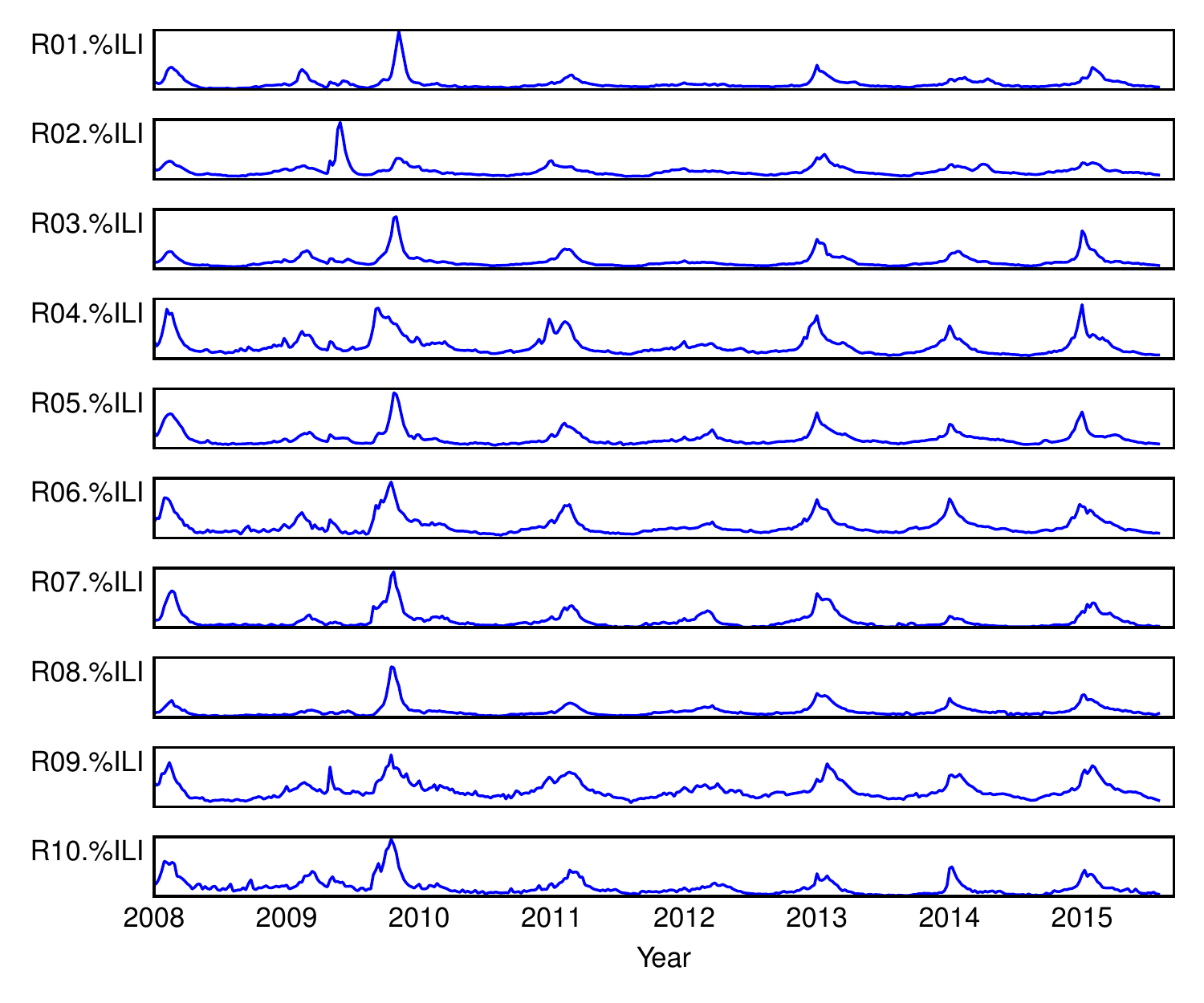}
\caption{Original flu time series}
\end{subfigure}%
\begin{subfigure}{.5\linewidth}
\captionsetup{skip=0pt}
\includegraphics[width=\textwidth]{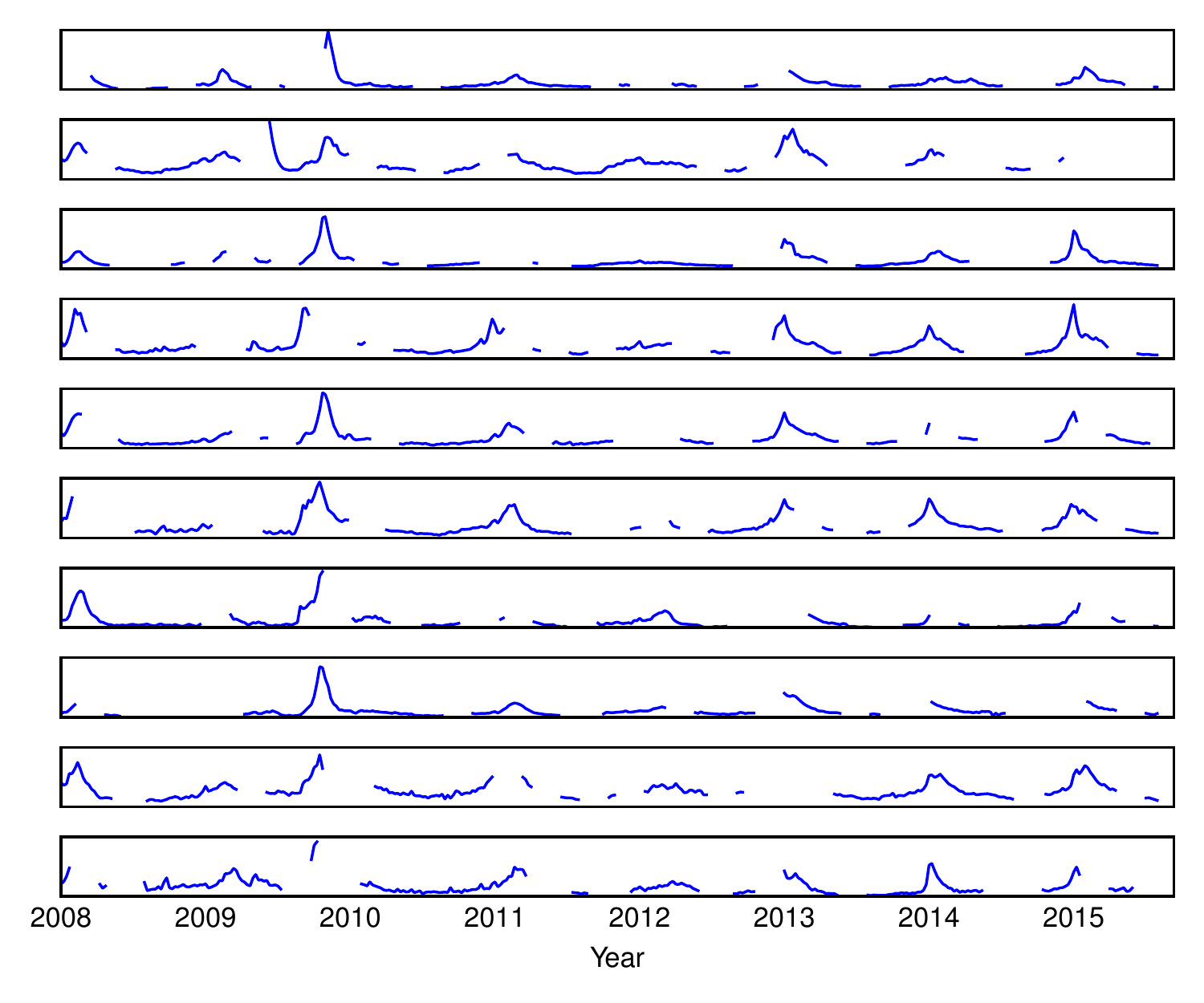}
\caption{Time series after dropping data}
\end{subfigure}
\begin{subfigure}{\linewidth}
\captionsetup{skip=0pt}
\includegraphics[width=\textwidth]{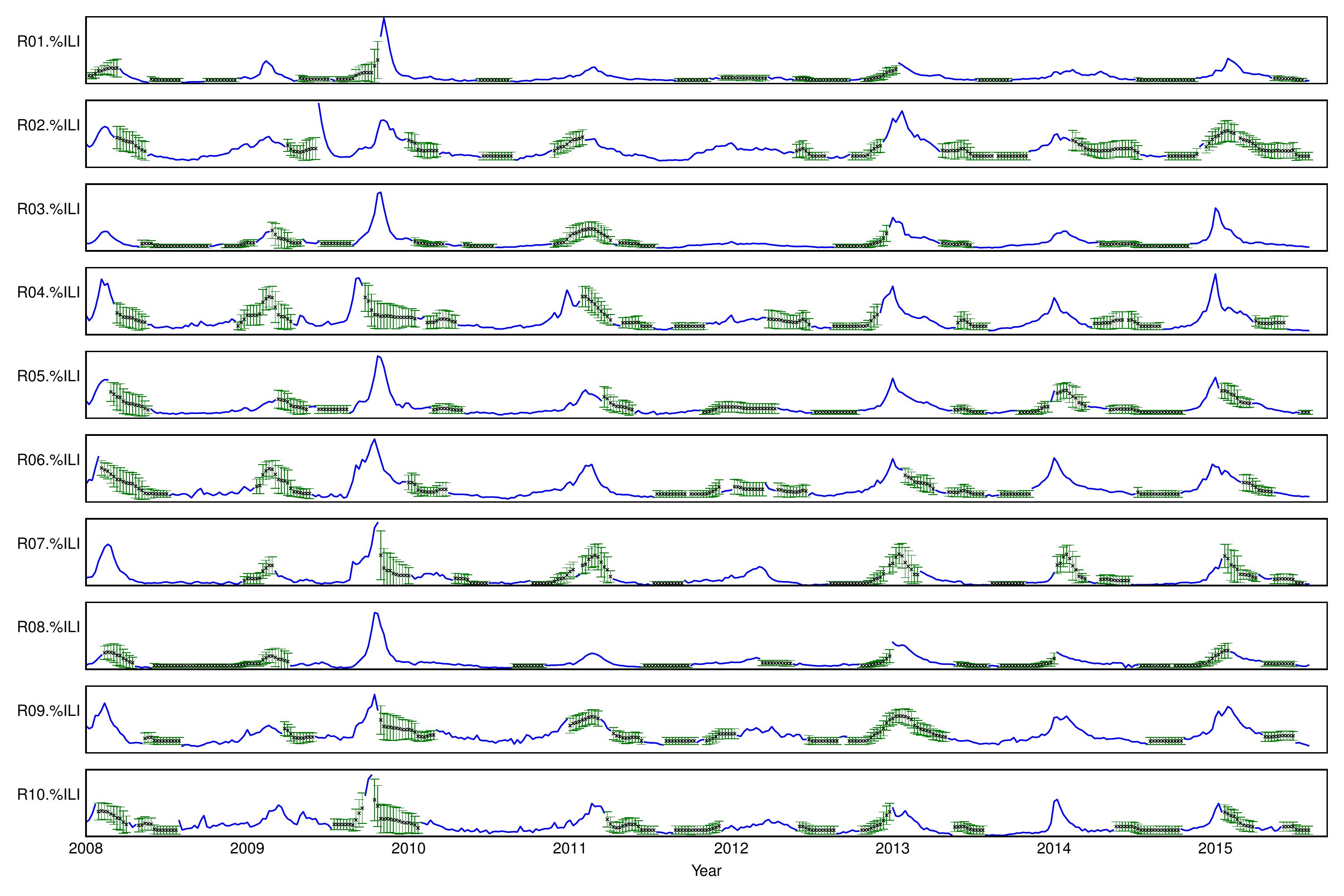}
\caption{Jointly imputed time series using TRCRP mixture ($p=10$)}
\label{subfig:observed-missing-data-all-imputed}
\end{subfigure}
\caption{Full, missing, and imputed flu time series over eight years in US
Regions 1 through 10.}
\label{fig:observed-missing-data-all}
\end{figure*}

The reason that small $p$ works well for jointly imputing the 10 time series in
Figure~\ref{fig:observed-missing-data-all} is that the multivariate TRCRP
mixture shares statistical strength across time series. Namely, when imputing a
missing value $x^{n_0}_t$ at time $t$ for time series $n_0$, the relevant
variables for predicting the hidden state $z_t$ are (i) the history
$\x^n_{t-p:t-1}$ of the current time series; and (ii) values $\set{x^{n}_{t}
\mid n_0 \ne n}$ of other time series at time $t$. The latter effect is the
dominant one in this imputation problem, leading to less sensitivity to $p$ than
might be expected.

\subsection{Clustering GDP time series}
\label{appx:experimental-methods-gapminder-clustering}

The clustering results from Figure~\ref{fig:cluster-datasets-gdp} were obtained
by using a TRCRP with a window of $p=5$ years. The nine clusters that are shown
were obtained by averaging dependence probabilities over $S=60$ posterior
samples (using a burn-in of 5000 iterations), and extracting groups of
variables whose dependence probabilities \eqref{eq:dependence-probability}
exceeded $80\%$. All time series in Figure~\ref{fig:cluster-datasets-gdp} are
linearly rescaled to $[0,1]$ for plotting purposes only.

\begin{figure*}[ht]
\begin{subfigure}{.5\linewidth}
\includegraphics[width=\linewidth]{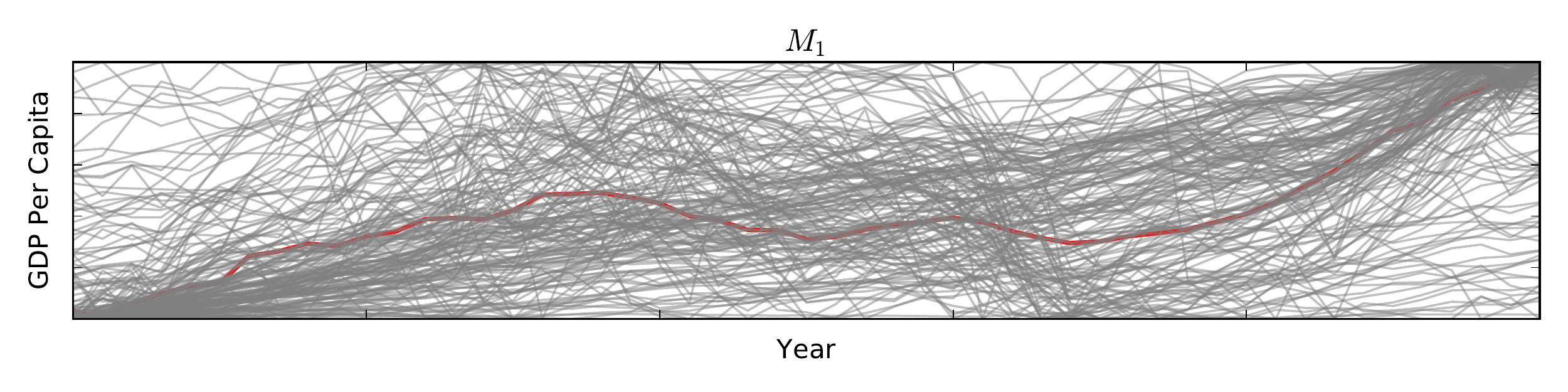}
\captionsetup{skip=0pt}
\subcaption{$k=1$}
\end{subfigure}
\begin{subfigure}{.5\linewidth}
\includegraphics[width=\linewidth]{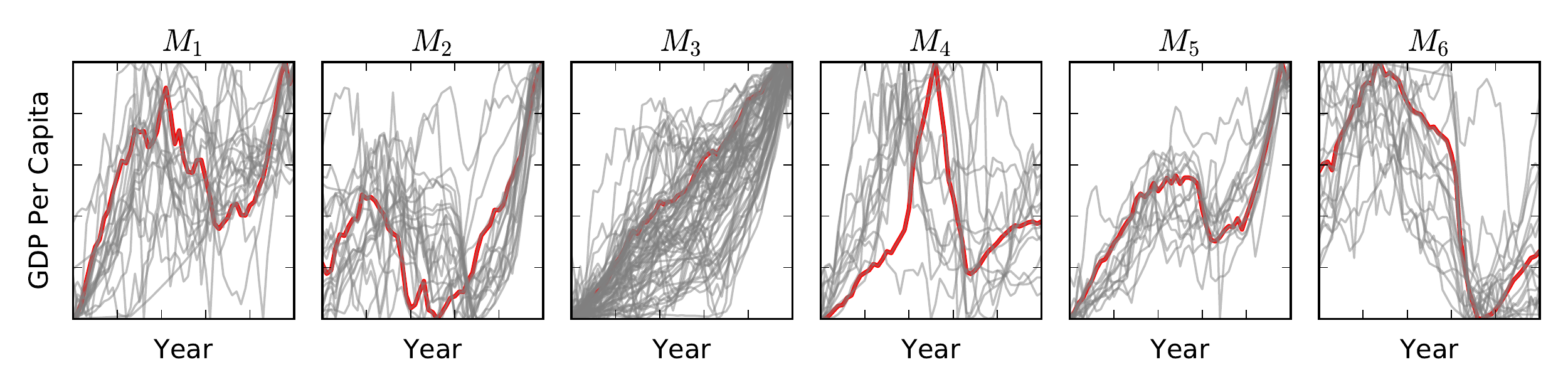}
\captionsetup{skip=0pt}
\subcaption{$k=6$}
\end{subfigure}
\begin{subfigure}{.5\linewidth}
\includegraphics[width=\linewidth]{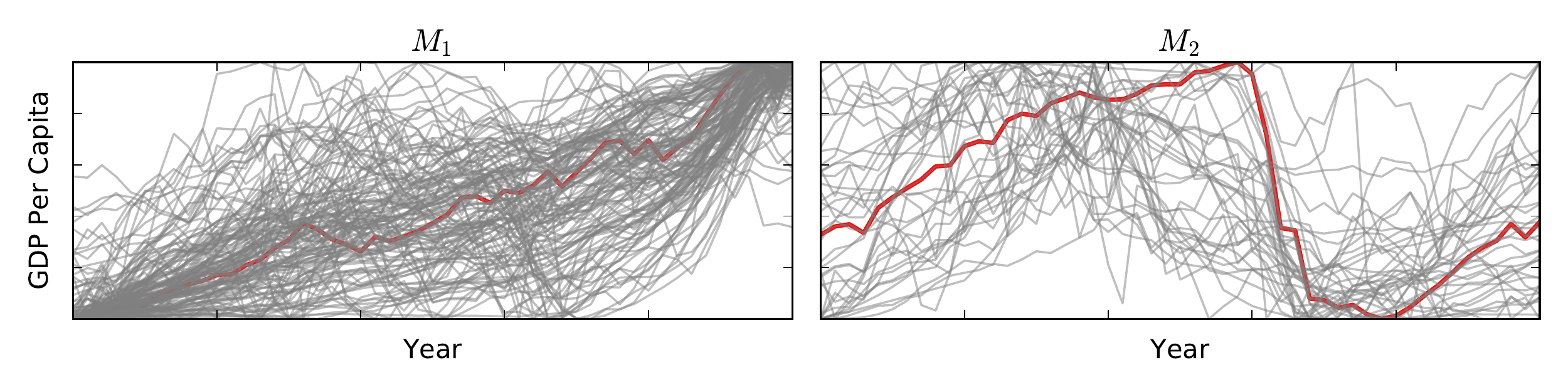}
\captionsetup{skip=0pt}
\subcaption{$k=2$}
\end{subfigure}
\begin{subfigure}{.5\linewidth}
\includegraphics[width=\linewidth]{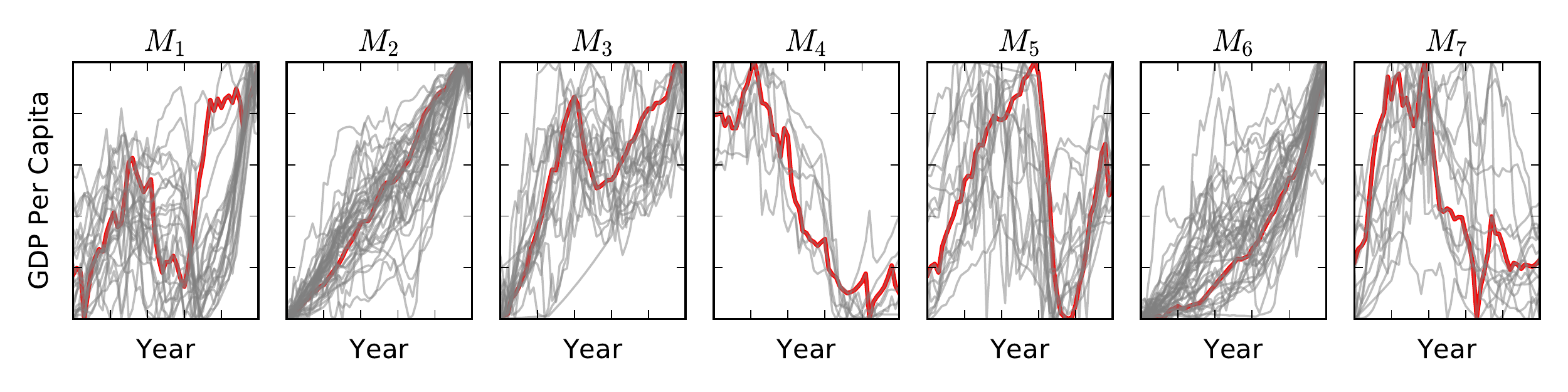}
\captionsetup{skip=0pt}
\subcaption{$k=7$}
\end{subfigure}
\begin{subfigure}{.5\linewidth}
\includegraphics[width=\linewidth]{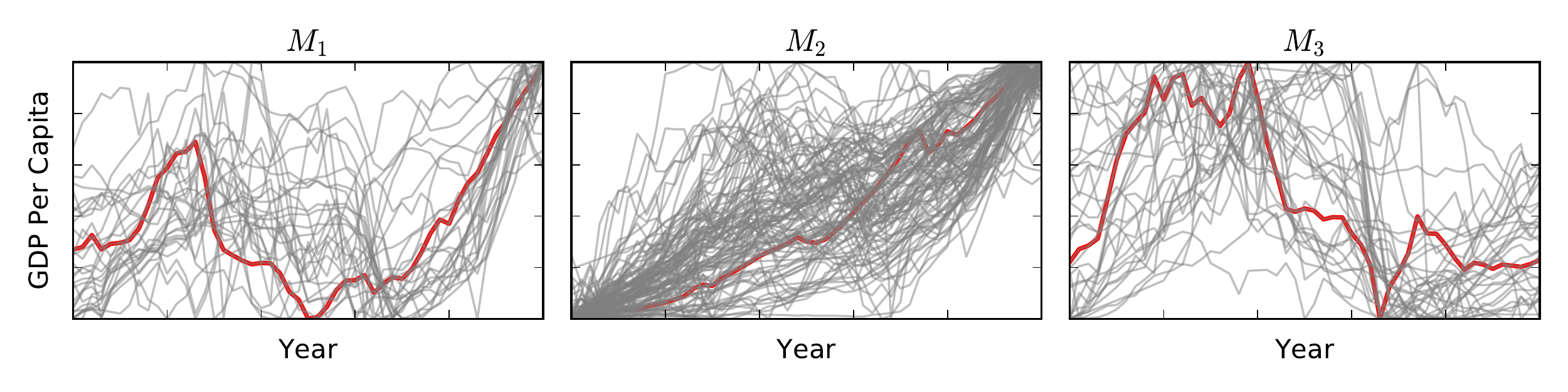}
\captionsetup{skip=0pt}
\subcaption{$k=3$}
\end{subfigure}
\begin{subfigure}{.5\linewidth}
\includegraphics[width=\linewidth]{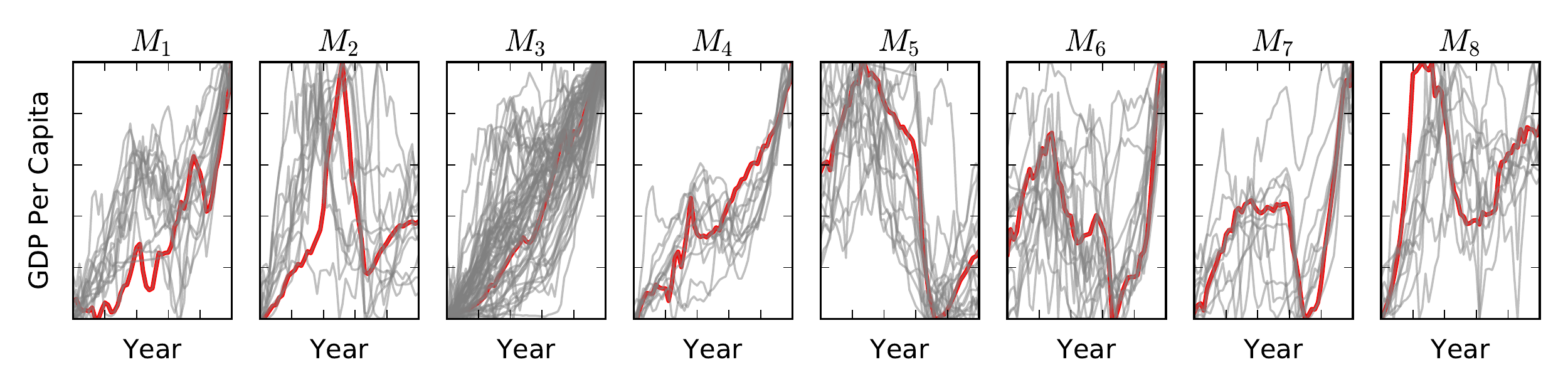}
\captionsetup{skip=0pt}
\subcaption{$k=8$}
\end{subfigure}
\begin{subfigure}{.5\linewidth}
\includegraphics[width=\linewidth]{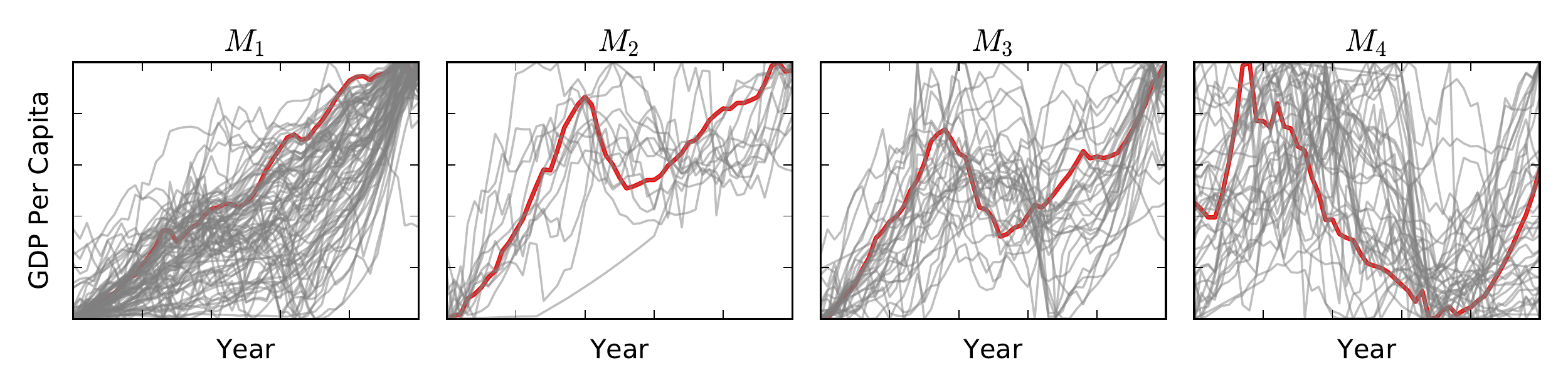}
\captionsetup{skip=0pt}
\subcaption{$k=4$}
\end{subfigure}
\begin{subfigure}{.5\linewidth}
\includegraphics[width=\linewidth]{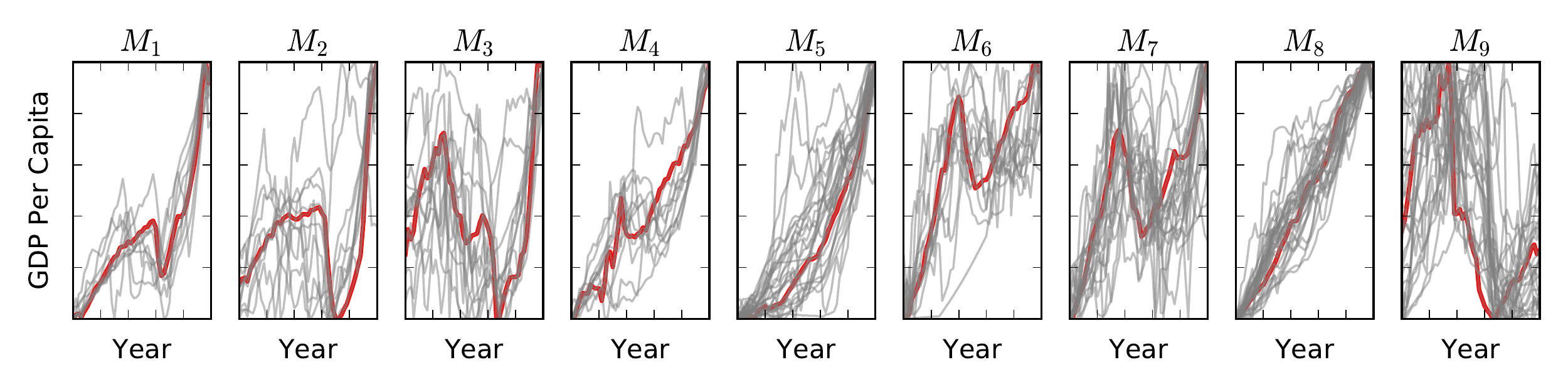}
\captionsetup{skip=0pt}
\subcaption{$k=9$}
\end{subfigure}
\begin{subfigure}{.5\linewidth}
\includegraphics[width=\linewidth]{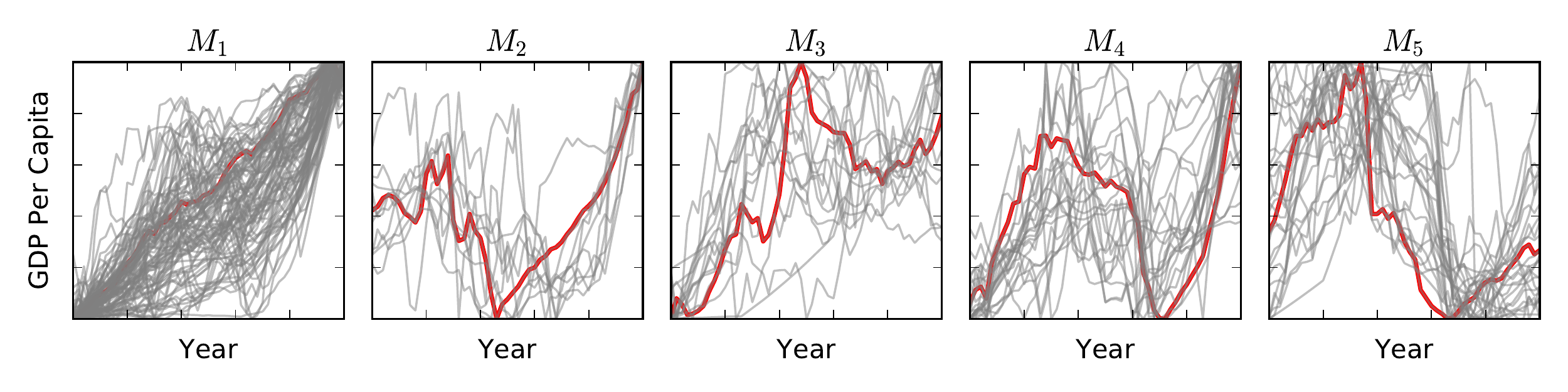}
\captionsetup{skip=0pt}
\subcaption{$k=5$}
\end{subfigure}
\begin{subfigure}{.5\linewidth}
\includegraphics[width=\linewidth]{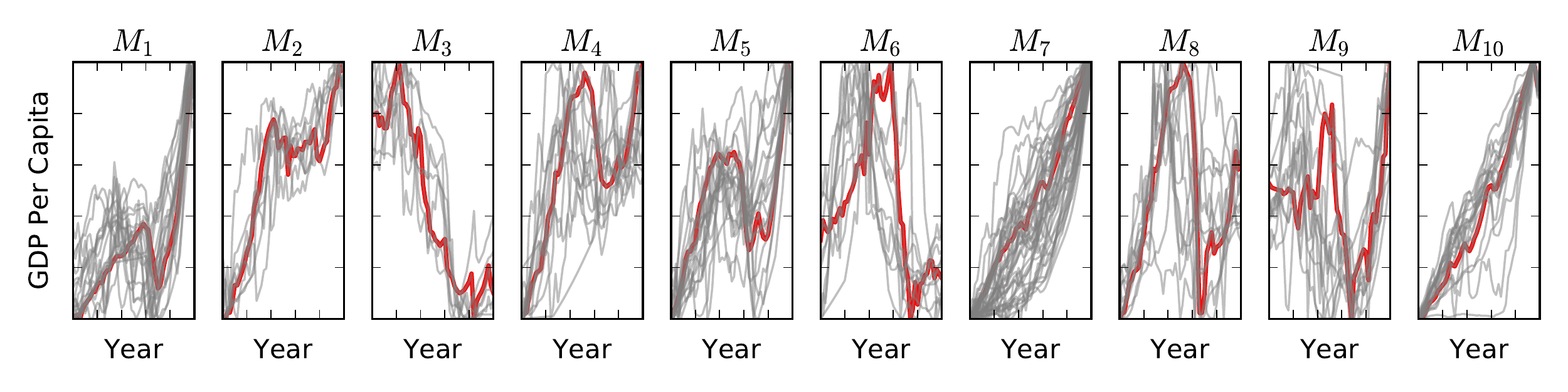}
\captionsetup{skip=0pt}
\subcaption{$k=10$}
\end{subfigure}

\caption{Outputs of k-medoids clustering on the GDP per capita time series for
all 170 countries in the Gapminder dataset, with $k=1,2,\dots,10$.
Distances are computed using the dynamic time warping (DTW) metric, a common
similarity measure between a pair of time series \citep{berndt1994}.
For each $k$, we randomly initialized the medoids and ran the algorithm to
convergence (medoids are shown in red, and time series assigned to that medoid
in gray).
Using k-medoids requires hand-tuning the number of latent clusters $k$, whereas
the proposed method (whose posterior clustering is shown in
Figure~\ref{fig:cluster-datasets-gdp} of the main text), places a non-parametric
Bayesian prior over this parameter.
Moreover, when compared to the clusters detected by the proposed method, those
detected by k-medoids with DTW appear qualitatively less distinct, and have more
repetitive and duplicated temporal patterns (especially apparent at higher $k$).
Finally, k-medoids outputs a fixed cluster assignment for each time series in
the population; these assignments are sensitive to the random initialization and
cannot be aggregated in a principled way.
In contrast, inference in the proposed method assigns probabilistic cluster
assignments that can be averaged coherently using
\eqref{eq:dependence-probability} to express posterior uncertainty.}
\label{fig:cluster-datasets-gdp-kmedoids}
\end{figure*}

While clustering is an unsupervised task that is challenging to evaluate
quantitatively (especially for real-world data, where there is no
``ground-truth''), qualitative comparisons to k-medoids clustering with the
dynamic time warping metric on the same GDP time series are shown and discussed
in Figure~\ref{fig:cluster-datasets-gdp-kmedoids}.

\subsection{Expanded results on clustering cell phone subscription time
series}
\label{appx:expanded-clustering}

\begin{figure*}
\centering

\begin{subfigure}{\linewidth}
\includegraphics[width=\linewidth]{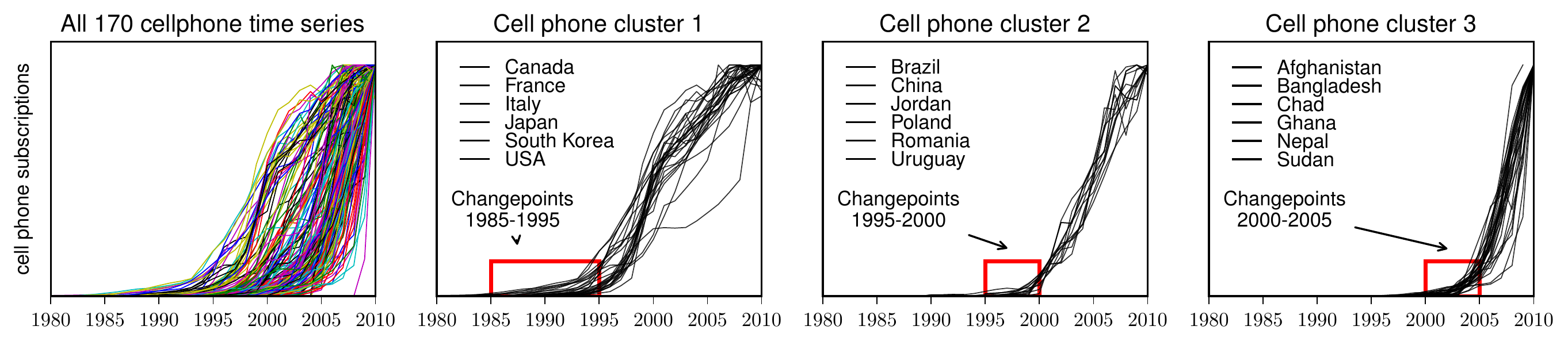}
\subcaption{Three posterior clusters in the TRCRP mixture correspond to three
non-overlapping change point windows.}
\label{subfig:cluster-datasets-cell-clusters}
\end{subfigure}
\bigskip
\bigskip

\begin{subfigure}{.5\linewidth}
\centering
\includegraphics[width=.8\linewidth]{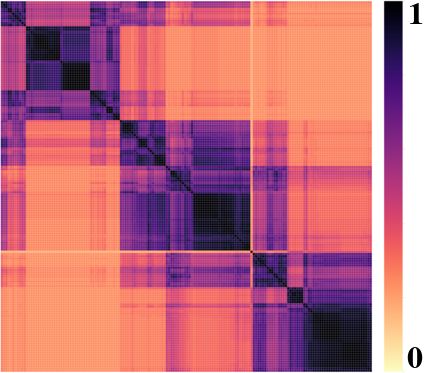}%
\subcaption{Pairwise dependence probability heatmap}
\label{subfig:cluster-datasets-cell-dependence-heatmap}
\end{subfigure}%
\begin{subfigure}{.5\linewidth}
\centering
\includegraphics[width=.8\linewidth]{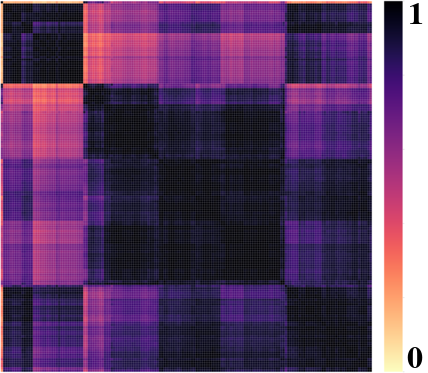}
\subcaption{Pairwise cross-correlation heatmap}
\label{subfig:cluster-datasets-cell-correlation-heatmap}
\end{subfigure}

\caption{Discovering changepoint patterns in cell phone subscriptions for 170
countries in the Gapminder dataset.
\subref{subfig:cluster-datasets-cell-clusters} The three clusters (extracted
from one posterior sample) correspond to three regimes each with non-overlapping
change point windows, annotated by red boxes.
The representative countries in each cluster have similar adoption times of cell
phone technology, a feature which differs across the clusters.
\subref{subfig:cluster-datasets-cell-dependence-heatmap} and
\subref{subfig:cluster-datasets-cell-correlation-heatmap} %
The matrix of dependence probabilities (averaged over 60 posterior samples using
\eqref{eq:dependence-probability}) and the matrix of pairwise
cross-correlations (bottom) between all pairs 170 time series. Each row and
column is a time series, and the color of a cell (a value between 0,1) indicates
the posterior dependence probability, resp. cross-correlation coefficient
(significant at the 0.05 level with Bonferroni correction). The TRCRP mixture
detects more refined dependence structures than those captured by linear
statistics.}
\label{fig:cluster-datasets-cell}
\end{figure*}

In addition to clustering GDP series from Figure~\ref{fig:cluster-datasets-gdp},
we applied the TRCRP prior with hierarchical extension
\eqref{eq:crp-over-time-series} to cluster historical cell phone subscription
data. The outcome of the clustering is shown in
Figure~\ref{fig:cluster-datasets-cell}, where we show all 170 time series in the
left most figure, along with three representative clusters from one posterior
sample. Each cluster corresponds to countries whose change point in cell phone
subscribers from zero to non-zero fell in a distinct window: 1985-1995 in
cluster 1, 1995-2000 in cluster 2, and 2000-2005 in cluster 3. We also compare
renderings of the the pairwise dependence probability matrix with the pairwise
cross-correlation matrix. Refer to the caption of
Figure~\ref{fig:cluster-datasets-cell} for additional details.

\end{appendices}

\end{document}